\documentclass[%
aps,prb,twocolumn,a4paper,showpacs,superscriptaddress,groupedaddress
]{revtex4-1}

\usepackage{braket}
\usepackage{graphicx}

\usepackage{hhline}
\usepackage{natbib}

\usepackage{amsmath}
\usepackage{amssymb}
\usepackage{bm}
\usepackage[usenames]{color}
\usepackage[normalem]{ulem}

\definecolor{myorange}{rgb}{0.7,0.5,0.0}
\definecolor{mygreen}{rgb}{0.0,0.7,0.0}
\definecolor{purple}{rgb}{0.75,0.0,1.0}
\definecolor{mahogany}{rgb}{0.8,0.0,0.0}

\newcommand{\tb}[1]{\textcolor{black}{#1}}

\newcommand{\tor}[1]{\textcolor{black}{#1}}

\newcommand{\txr}[1]{\textcolor{black}{#1}}

\newcommand{\tbs}[1]{\textcolor{blue}{\def\ULthickness{1pt}\sout{#1}}}

\newcommand{\mi}[1]{\textcolor{black}{#1}}
\newcommand{\mib}[1]{\textcolor{black}{#1}}

\newcommand{\mts}[1]{\textcolor{black}{#1}}

\newcommand{\sP}{\mathcal{P}}
\newcommand{\sL}{\mathcal{L}}

\newcommand{\ra}{\rangle}

\newcommand{\Ns}{N_{\text{s}}}

\begin{document}
  \title{{\textit{Ab initio} material design of Ag-based oxides for high-$T_c$ superconductor}} 
  \author{Motoaki Hirayama$^{1,2,3)}$\email{hirayama@ap.t.u-tokyo.ac.jp}, Michael Thobias Schmid $^{4)}$,  Terumasa Tadano$^{5)}$, Takahiro Misawa$^{6)}$, and Masatoshi Imada$^{4,7)}$}
  \affiliation{$^{1)}$Department of Applied Physics, University of Tokyo, 7-3-1 Hongo, Bunkyo-ku, Tokyo 113-8656, Japan}
  \affiliation{$^{2)}$RIKEN Center for Emergent Matter Science, Wako, Saitama 351-0198, Japan}
  \affiliation{$^{3)}$JST, PRESTO, Hongo, Bunkyo-ku, Tokyo 113-8656, Japan}
   \affiliation{$^{4)}$Waseda Research Institute for Science and Engineering, Waseda University, Shinjuku, Tokyo 169-8555, Japan}
  \affiliation{$^{5)}$Research Center for Magnetic and Spintronic Materials, National Institute for Materials Science, Tsukuba 305-0047, Japan}   
  \affiliation{$^{6)}$Beijing Academy of Quantum Information Sciences, Haidian District, Beijing 100193, China}
  \affiliation{$^{7)}$Toyota Physical and Chemical Research Institute, Nagakute, Aichi 480-1192, Japan} 
  
\begin{abstract}
We propose silver-based oxides with layered perovskite structure as candidates of exhibiting intriguing feature of strongly correlated electrons. 
The compounds show unique covalence between Ag $d$ and O $p$ orbitals with \mib{the} strong electron correlation of their antibonding orbital similar to the copper oxide \mib{high-temperature superconductors}, but stronger covalency and slightly smaller correlation strength. 
We examine $A_2$AgO$_2$X$_2$ with $A=$Sr and Ba and $X=$F, I and Cl in detail. \mib{Among them, Sr$_2$AgO$_2$F$_2$ has the largest effective onsite Coulomb repulsion and} shows \mi{an antiferromagnetic insulating ground state, which  competes with 
correlated metals. 
It offers features both similar and distinct from the copper oxides, \mib{and paves a new route}. The possibility of superconductivity in doped systems is discussed.} 
\end{abstract}

\maketitle

\section{Introduction}
Copper oxides with layered perovskite structure (abbreviated as cuprates hereafter)~\cite{bednortz} exhibit spectacular properties with high-temperature superconducting phases above 100K, which is the highest record so far at ambient pressure. Above the superconducting critical temperature, they show enigmatic pseudogap and bad metal behaviors. Recent {\it ab initio} calculations of several cuprate superconductors~\cite{hirayama18,hirayama19,ohgoe20} suggest that the superconductivity is hampered  by electronic charge instability actually observed experimentally such as stripe order~\cite{tranquada} or mesoscopic-scale phase separation~\cite{pan}, when the correlation strength is too high, which yields emergent too strong effective attraction. \mib{The optimum combination of onsite and off-site correlation strengths and band structure to enhance and stabilize superconductivity is not clarified yet and it could be out of the accessible range of the cuprate parameters} in general. 
In addition, the level difference between the copper $d$ and the oxygen $p$ orbitals is roughly fixed to be 3 eV, while this difference may also be an important parameter to optimize the superconductivity and is desired to \tor{be controlled}. 
 
In comparison to the 3$d$ transition metal elements, to which copper belongs, the 4$d$ and 5$d$ transition metal elements have larger radius of $d$ orbitals, which makes the onsite Coulomb interaction smaller, while at the same time makes the overlap of neighboring orbitals larger for similar lattice constant. Furthermore, expected smaller level difference between Ag 4$d$ (Au $5d$) and O 2$p$ orbitals than the cuprates makes larger covalency and the larger Wannier spread of the antibonding orbital that makes the interaction strength smaller and the transfer larger. If we can design the block layer having the weak screening effect, we will be able to create a system that not only has large hopping but also strong interaction. Therefore, it is desired to pursue the 4$d^9$ and $5d^9$ compounds on the basis of {\it ab initio} calculations to predict the feature of electron correlation also as a platform of possible superconducting phase. 

\begin{figure}[htp]
	\includegraphics[clip,width=0.25\textwidth ]{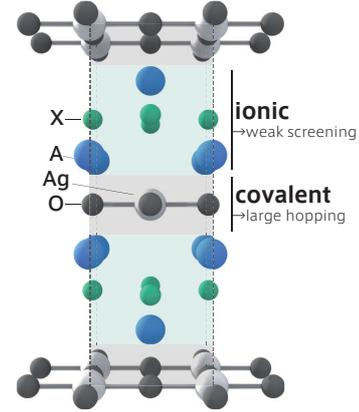}
	\caption{(Color online)
		Crystal structure of $A_2$AgO$_2X_2$ ($A=$Sr and Ba, $X$=F, Cl, Br, and I), which belongs to T'-type structure studied in this paper.
	}
	\label{Sr2AgO2F2_structure}
\end{figure}
 
\begin{figure}[htp]
	\includegraphics[clip,width=0.4\textwidth ]{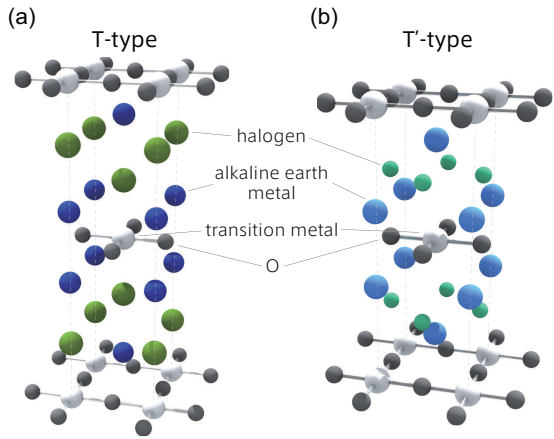}
	\caption{(Color online)
		(a) La$_2$CuO$_4$-type (T-type) structure.
		(b) Nd$_2$CuO$_4$-type (T'-type) structure. 
	}
	\label{str}
\end{figure}

\begin{figure}[htp]
	\centering 
	\includegraphics[clip,width=0.27\textwidth ]{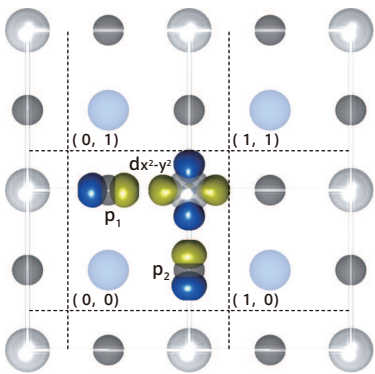} 
	\caption{(Color online) 
		Position of the $d_{x^2-y^2}$ and O $2p$ orbitals 
		and definition of the $2\times 2$ sublattice used in the VMC for the three-band Hamiltonian.
		Phases of the orbitals are distinguished by yellow ($+$) and blue ($-$) colors.
	}
	\label{dpp2x2}
\end{figure} 

In this paper we propose several silver as well as gold oxide compounds illustrated by typical crystal structure in Fig.~\ref{Sr2AgO2F2_structure} and derive their {\it ab initio} effective Hamiltonians after structural optimization and confirmation of the material stability to be used to predict electronic properties including correlation effects and superconductivity. \mi{To show its promising feature, we supplement with an attempt to solve one of the Hamiltonians for Sr$_2$AgO$_2$F$_2$, which has the largest Coulomb interaction relative to the nearest-neighbor transfer integral, by using a quantum many-body solver based on the variational Monte Carlo (VMC) method. It indeed shows that the mother compound has} \mts{an} \mi{antiferromagnetic Mott insulating ground state competing with} \mts{a correlated metallic state.}
The structures we study is essentially T-type and T'-type layered perovskite structure illustrated in Figs.~\ref{str}(a) and (b). 
The common lattice structure consisting of the transition metal $d$ and oxygen $p$ orbitals in our compounds are sketched in Fig.~\ref{dpp2x2}.

The organization of the paper is the following:
In Sec.~\ref{Method}, we briefly describe our {\it ab initio} method and computational conditions. We then propose our strategy of materials design to find promising strongly correlated systems that are distinct from the cuprates in Sec.~\ref{Strategy}. In Sec.~\ref{Result}, the procedure of deriving Hamiltonians and the obtained effective Hamiltonians are presented for $A_2$AgO$_2X_2$, and  $A_2$AuO$_2X_2$, where $A=$Sr and Ba, and $X=$F, Cl, Br, and I after structural optimization. A preliminary result of the VMC calculation for Sr$_2$AgO$_2$F$_2$ using the derived Hamiltonian is also shown in Sec.~\ref{Result} as a test case, which shows a rich and promising feature. Details of the VMC results will be reported elsewhere. Section \ref{Concluding} is devoted to concluding remark.

\begin{figure*}[htp]
	\includegraphics[clip,width=0.5\textwidth ]{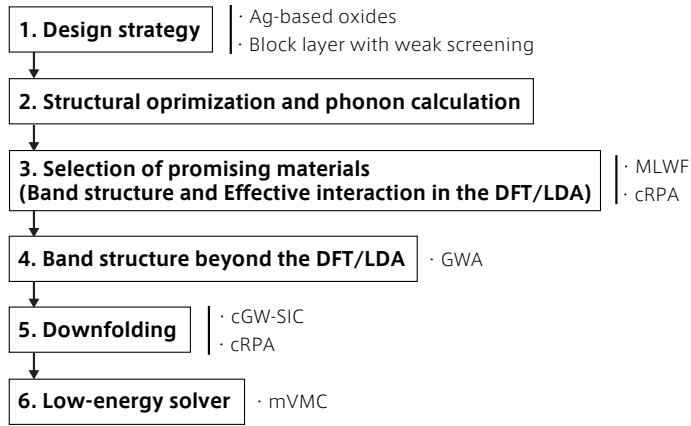}
	\caption{(Color online) 
		Overview of material design scheme for strongly correlated systems in this paper. 
		We focus on 
       Ag- and Au-based compounds. MLWF, cRPA, GWA, cGW-SIC, and mVMC 
       stand for maximally localized Wannier function, constrained random phase approximation, GW approximation, constrained GW approximation supplemented by the self-interaction correction, and many-variable variational Monte Carlo,
       respectively. See Sec.~\ref{Method} for the details.
	}
	\label{outline}
\end{figure*}

\section{Method} \label{Method}
Strongly correlated electron systems have been a challenging field for theoretical understanding for decades\cite{imada1998RMP}. In particular,  density functional calculation widely applied to predict electronic properties does not offer a satisfactory scheme because of the failure of representing strong correlation phenomena such as the Mott insulator~\cite{imada1998RMP}. To overcome the difficulty, by taking account of the universal hierarchical nature of the strongly correlated systems, which necessarily consist of the sparse and simple bands near the Fermi level, an efficient scheme based on the spirit of the renormalization group has been developed to derive effective Hamiltonians for such low-energy simple bands called the ``target bands". We apply this multi-scale {\it ab initio} scheme for correlated electrons (MACE)~\cite{imadamiyake10} to derive low-energy effective Hamiltonian for the purpose of enabling the next step at which the derived Hamiltonian is solved by accurate quantum many-body solvers and to quantitatively predict physical quantities including superconductivity.
This procedure is embedded in the basic flow chart of our materials design sketched in Fig.~\ref{outline}.

In the first step, we design the structure of the Ag- and Au-based oxides in terms of energy scales and feasible stability of the crystal structure. 
Next, we perform structural optimization for the candidate materials and calculate the phonon spectra to see the stability of the structure. By using the optimized structure, we derive the electronic effective Hamiltonian for the degrees of freedom near the Fermi level. We derive three-band Hamiltonians called $dpp$ Hamiltonian consisting of Ag 4$d_{x^2-y^2}$ (or Au 5$d_{x^2-y^2}$) and two O 2$p$ orbitals as well as one-band Hamiltonian for the antibonding band of these three orbitals. The forms of the effective Hamiltonians themselves already allow us to infer the correlation effect within the scope of this paper by comparing the {\it ab initio} Hamiltonian parameters with those of other existing materials such as the cuprates, which were already derived in a few compounds~\cite{hirayama18,hirayama19}. We also supplement our proposal by solving the derived Hamiltonian by the VMC method, which indeed shows antiferromagnetic insulating phase for the mother compound, but with smaller Mott gap than the cuprates, which anticipates intriguing properties upon carrier doping. The VMC method is summarized in Appendix \ref{VMCmethod}.

\subsection{Downfolding method}
In this subsection, we briefly summarize the method of deriving the effective Hamiltonian by taking partial trace summation of the high-energy degrees of freedom far from the Fermi level.
See Refs.~\onlinecite{hirayama18} and \onlinecite{hirayama19} for detailed procedure. The effective Hamiltonian in the low-energy space (target space) is 
given in the form of extended Hubbard-type Hamiltonian
without any adjustable parameters as
\begin{eqnarray}
\mathcal{H}_{\text{eff}} ^{\text{cGW-SIC}}= \sum_{ij} \sum_{\ell_1 \ell_2\sigma }&&
t^{\text{cGW-SIC}}_{\ell_1 \ell_2\sigma}(\bm{R}_i-\bm{R}_j) d_{i\ell_1\sigma} ^{\dagger} d_{j \ell_2\sigma} \nonumber \\
+ \frac{1}{2} \sum_{i_1i_2i_3i_4} \sum_{\txr{\ell_1 \ell_2 \ell_3 \ell_4} \sigma \eta \rho \tau}  
&\biggl\{& W_{ \ell_1 \ell_2 \ell_3 \ell_4\sigma \eta \rho \tau }^r(\bm{R}_{i_1},\bm{R}_{i_2},\bm{R}_{i_3},\bm{R}_{i_4}) \nonumber \\
&&d_{i_1 \ell_1\sigma}^{\dagger}d_{i_2 \ell_2\eta} d_{i_3 \ell_3\rho}^{\dagger} d_{i_4 \ell_4\tau}\biggl\},
\label{Hamiltonian0}
\end{eqnarray}
where $d_{i\ell\sigma} ^{\dagger}$ ($d_{i \ell\sigma}$) is the creation
(annihilation) operator of an electron for the $\ell$th \tor{maximally} localized Wannier function (MLWF) with spin $\sigma$ centered at unit cell $\bm{R}_i$.
Here, the single-particle term 
\begin{equation}
t^{\text{cGW-SIC}}_{ \ell_1 \ell_2\sigma}(\bm{R})= \langle \phi _{ \ell_1\bm{0}}|{H}^{\text{cGW-SIC}}_{K}|\phi _{ \ell_2\bm{R}} \rangle, 
\label{cGW-SICK}
\end{equation}
and  the interaction term 
\begin{eqnarray}
W_{ \ell_1 \ell_2 \ell_3 \ell_4\sigma \eta \rho \tau }^r(\bm{R}_{i_1},\bm{R}_{i_2},\bm{R}_{i_3},\bm{R}_{i_4}) \nonumber \\
= \langle \phi _{ \ell_1\bm{R}_{i_1}}\phi _{ \ell_2\bm{R}_{i_2}}|{H}^{\text{cGW-SIC}}_{W^r}|\phi _{ \ell_3\bm{R}_{i_3}}\phi _{ \ell_4\bm{R}_{i_4}} \rangle ,
\label{cGW-SICW}
\end{eqnarray}
will be derived in this paper, where $\phi _{ \ell\bm{R}_{i}}$ is the $\ell$th MLWF centered at $\bm{R}_i$.
Following Refs.~\onlinecite{hirayama18} and \onlinecite{hirayama19}, the one-body term 
(\ref{cGW-SICK})
and the two-body term (\ref{cGW-SICW}) 
were calculated by the 
constrained GW approximation (cGW) with the self-interaction correction (SIC) ~\cite{hirayama13} and 
the constrained random phase approximation (cRPA)~\cite{aryasetiawan04}, 
respectively, using the Green's function of the band structure of all degrees of freedom.
It should be noted that all the parameters in 
Eq.(\ref{Hamiltonian0}), namely $t^{\text{cGW-SIC}}$ and $W^r$, are given 
from the first principles calculation without any adjustable parameters. 

In the later presentation, we often use the notation $U(0,0,0)$ or $U$, $V(l,m,n)$ and $t(l,m,n)$ for important parameters. These denote $(\ell_1,\ell_2)$ component of onsite interaction matrix $U_{\ell_1,\ell_2}=W^r_{\ell_1,\ell_1,\ell_2,\ell_2,\sigma_1,\sigma_1,\sigma_2,\sigma_2,}(\bm{R}_{0},\bm{R}_{0},\bm{R}_{0},\bm{R}_{0})$,
intersite Coulomb repulsion matrix $V_{\ell_1,\ell_2}(l,m,n)=W^r_{\ell_1,\ell_1,\ell_2,\ell_2,\sigma_1,\sigma_1,\sigma_2,\sigma_2,}(\bm{R}_{0},\bm{R}_{0},\bm{R}_{0}+\bm{R},\bm{R}_{0}+\bm{R})$ and transfer $t_{\ell_1,\ell_2}(l,m,n)=t_{\ell_1, \ell_2, \sigma}(\bm{R})$  where $\bm{R}=(l,m,n)$ in the length unit of the unitcell.

In this research, we follow the basic strategy of MACE. We first derive effective Hamiltonians based on the local density approximation (LDA) of the density functional theory (DFT) to capture the over all trend and to save the computational load.  
However, we next follow the same way as Refs.~\onlinecite{hirayama18} and \onlinecite{hirayama19} and use the whole band structure obtained by the GW approximation (GWA) beyond the DFT/LDA
to derive the most elaborated {\it ab initio} Hamiltonian for some Ag compounds.

In the cGW\cite{hirayama13,hirayama17}, the band dispersion is determined 
from the self-energy and the polarization by excluding the 
contribution from the low-energy degrees of freedom to 
remove the double counting; 
The cGW method can explicitly exclude the double counting of
the exchange correlation energy in the effective Hamiltonian.
In this paper, we further take into account the chemical potential correction to make the orbital filling fixed at the GW result by following Ref.~\onlinecite{hirayama19}.
For details of the method see Ref.~\onlinecite{hirayama19}.
On the computational conditions related to phonon, DFT and GW, and VMC calculations, see Appendix~\ref{CompCond}.

\section{Design strategy of ${\bf Ag}$-based superconductors} \label{Strategy}

An important quantity throughout this paper is the effective correlation strength $|U/t|$ characterized by the ratio of the effective onsite interaction $U$ and the hopping amplitude $t$ between the nearest neighbor orbitals in the one-band effective Hamiltonian, which we will derive below. The intersite interaction $V_1$ (for the nearest neighbor) and $V_2$ (for the next nearest neighbor) relative to $U$ will also turn out to be important.
Here, we summarize our idea of materials design in this paper by referring to these parameters. 
We propose a new silver compound with (AgO$_2$)$^{-2}$ plane as candidates showing promising feature by using the {\it ab initio} method.

Ag has an electronegativity closer to that of O than to that of Cu.
Therefore, Ag and O in the AgO$_2$ plane would form a strong covalent bond and would have a larger energy scale hopping than in the case of the CuO$_2$ plane.
However, the bare Coulomb interaction of the Ag $4d$ orbital is anticipated to be weaker than that of the Cu $3d$ orbital, and the correlation of the Ag-based system becomes weaker.

\begin{figure*}[htp]
	\includegraphics[clip,width=0.55\textwidth ]{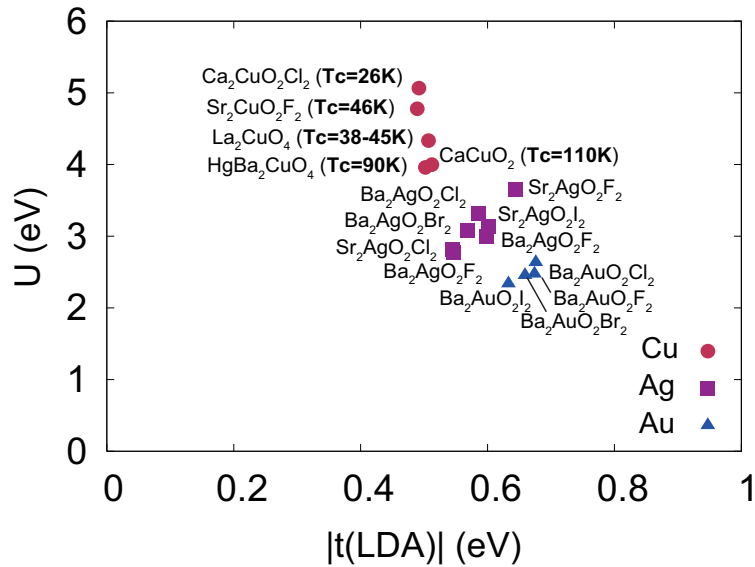}
	\caption{(Color online)
	Layered perovskites with Cu, Ag and Au are placed in the parameter plane of the nearest neighbor hopping $t$ in the LDA level and the on-site effective interaction $U$
\tor{for the single-band Hamiltonian}
in the Cu-, Ag-, and Au-based superconductors
\tor{calculated by the cRPA method from the DFT band structures, which corresponds to the calculation in Section IV. B}.
Plots of Ag and Au compounds are one of the main results in the present paper.
	}
	\label{tU_LDA}
\end{figure*}
As shown in Fig.~\ref{tU_LDA}, it is observed that the smaller the $U$ value (or equally $|U/t|$ because $t$ is similar), the higher the $T_c$ tends to be for the known copper oxides. This trend is consistent with our former analyses that too strong $|U/t|$ generates too strong effective attraction of carriers, which eventually results in the charge inhomogeneity instead of the superconductivity~\cite{misawaHubbard,misawainterface,imada_suzuki,imadareview21}.
We note that this trend is opposite to the observation in the literature~\cite{nilsson}.
\mib{In fact, it was pointed out that larger $|U/t|$ tends to show higher $T_c$ in the comparison of Bi$_2$Sr$_2$CuO$_6$, Bi$_2$Sr$_2$CaCuO$_8$, HgBa$_2$CuO$_4$ and CaCuO$_2$~\cite{Moree2022}. This implies that there exists the optimum value of $|U/t|$ to maximize $T_c$. In addition, it was shown that the off-site interactions have substantial effects on the amplitude of the superconducting order parameter. The amplitude also quantitatively depends on the detailed band structure. Therefore, it is desired to further cultivate variety in different family of \tor{compounds} by materials design to reach comprehensive and quantitative understanding of the superconductivity, which also contributes to identify the superconducting mechanism.}

\mib{Since Ag has a smaller bare interaction than the cuprates as we will show later, it offers an unprecedented approach from the weaker correlation side to reach the optimum}: We try to enlarge the correlation effect of Ag-based compounds by designing the block layer to make weak screening to reach the optimum correlation and to achieve high $T_c$.
Block layers containing halogens fit for this purpose by suppressing the screening:
The closed $p$-orbital band of halogen is stable and appears far from the Fermi level, so the screening is expected to be weaker.
In fact, the halogen system of copper oxides has a large $U$ as we will show later.

Halogens are also useful in terms of the valence of Ag.
Ag is less likely to have an oxidation number greater than $+1$, which is smaller than Cu.
However, by using a strong anion such as fluorine, it is possible to obtain a valence other than $+1$.
For example, Ag in Cs$_2$AgF$_4$ is a divalent anion and has the same La$_2$CuO$_4$ structure (T structure) as K$_2$CuF$_4$.

In this paper, we consider $A_2$AgO$_2X_2$ (A=Ca,Sr,Ba, X=F,Cl,Br,I) with T and T' structures.
These Ag compounds have not been reported in experiments yet.
However, there exist similar mixed-anion systems~\cite{Kageyama2018} in case of copper such as Ca$_2$CuO$_2$Cl$_2$ and Sr$_2$CuO$_2$F$_2$ and palladium such as Ba$_2$PdO$_2$F$_2$ and Ba$_2$PdO$_2$Cl$_2$.
We also calculate Au-based oxides with a similar block layer for comparison.
The Au-based oxide is expected to have an even smaller correlation than that of Ag.

\section{Result} \label{Result}
We here present the {\it ab initio} low-energy Hamiltonians for $A_2$AgO$_2X_2$ and 
$A_2$AuO$_2X_2$ with $A=$Sr and Ba and $X=$F, Cl, Br and I. We also discuss basic properties for Sr$_2$AgO$_2$F$_2$ by solving the {\it ab initio} Hamiltonian using
the mVMC. 

\subsection{Dynamical stability}

\begin{table}[htb]
	\caption{Structural parameters of the studied T'-type $A_{2}$AgO$_{2}X_{2}$ (space group: $I4/mmm$) optimized with PBEsol functional. The $A$ atoms occupy the 4e Wyckoff site $(0,0,\pm z_A)$. The Ag, O, and $X$ atoms are located at the 2a, 4c, and 4d sites, respectively.}
	\label{table:structures_Tprime_Ag}
	\begin{ruledtabular}
		\begin{tabular}{lcccc}
			Composition & $a$ (\AA) & $c$ (\AA) & $c/a$ & $z_A$ \\
			\hline
			Sr$_{2}$AgO$_{2}$F$_2$ &  4.146 & 12.428 & 2.9974 & 0.3632 \\
			Sr$_{2}$AgO$_{2}$Cl$_2$ &  4.359 & 14.083 & 3.2306 & 0.3844 \\
			Sr$_{2}$AgO$_{2}$Br$_2$ &  4.439 & 14.759 & 3.3247 & 0.3942 \\
			Sr$_{2}$AgO$_{2}$I$_2$ &  4.585 & 15.670 & 3.4174 & 0.4056 \\
			Ba$_{2}$AgO$_{2}$F$_2$ &  4.268 & 13.668 & 3.2028 & 0.3649 \\
			Ba$_{2}$AgO$_{2}$Cl$_2$ &  4.467 & 15.254 & 3.4145 & 0.3857 \\
			Ba$_{2}$AgO$_{2}$Br$_2$ &  4.536 & 15.854 & 3.4952 & 0.3947 \\
			Ba$_{2}$AgO$_{2}$I$_2$ &  4.665 & 16.627 & 3.5640 & 0.4047 \\
		\end{tabular}
	\end{ruledtabular}
\end{table}
\begin{table}[htb]
	\caption{Structural parameters of the studied T-type $A_{2}$AgO$_{2}X_{2}$ (space group: $I4/mmm$) optimized with PBEsol functional. The $A$ and $X$ atoms occupy the 4e site $(0,0,\pm z)$, and Ag atoms and O atoms are located at the 2a and 4c sites, respectively.}
	\label{table:structures_T_Ag}
	\begin{ruledtabular}
		\begin{tabular}{lccccc}
			Composition & $a$ (\AA) & $c$ (\AA) & $c/a$ & $z_A$ & $z_X$ \\
			\hline
			Sr$_{2}$AgO$_{2}$F$_2$ &  4.093 & 13.293 & 3.248 & 0.3749 & 0.1970 \\
			Sr$_{2}$AgO$_{2}$Cl$_2$ &  4.155 & 15.245 & 3.669 & 0.3900 & 0.1884 \\
			Sr$_{2}$AgO$_{2}$Br$_2$ &  4.184 & 16.245 & 3.883 & 0.3974 & 0.1851 \\
			Sr$_{2}$AgO$_{2}$I$_2$ &  4.215 & 18.489 & 4.386 & 0.4116 & 0.1760 \\
			Ba$_{2}$AgO$_{2}$F$_2$ &  4.145 & 14.323 & 3.455 & 0.3721 & 0.1895 \\
			Ba$_{2}$AgO$_{2}$Cl$_2$ &  4.258 & 16.148 & 3.792 & 0.3878 & 0.1857 \\
			Ba$_{2}$AgO$_{2}$Br$_2$ &  4.303 & 16.932 & 3.935 & 0.3941 & 0.1852 \\
			Ba$_{2}$AgO$_{2}$I$_2$ &  4.359 & 18.478 & 4.240 & 0.4050 & 0.1821 \\
		\end{tabular}
	\end{ruledtabular}
\end{table}
The structural parameters of the designed Ag-based oxides optimized by using the PBEsol functional are summarized in Tables \ref{table:structures_Tprime_Ag} and \ref{table:structures_T_Ag}. As is shown in Fig.~\ref{Sr2AgO2F2_structure}, two halogen ions in the T'-type structure are located in the same $X_2$ plane, whereas they are in different but adjacent planes in the T-type structure, where T and T' structures are illustrated in Fig.~\ref{str}. Hence, the in-plane (out-of-plane) lattice constant of the T'-type structure becomes larger (smaller) than that of the T-type. The relative stability of the two structures is also evaluated from the difference of the ground state energy; the T'-type structure is energetically more stable than the T-type only for $X=$ F, whereas the T-type is preferable for larger halogen ions ($X=$ Cl, Br, and I). This tendency is consistent with the experimental findings on the isostructural Cu oxides; Sr$_2$CuO$_2$F$_2$ displays the T'-type structure~\cite{Kissick}, while Sr$_2$CuO$_2$Cl$_2$ crystallizes in the T-type~\cite{Miller}. The same tendency was also reported in a recent theoretical study of nickel oxides~\cite{Hirayama_materials_design_prb2020}.

\begin{figure*}[htp]
	\centering
	\includegraphics[width=0.8\textwidth,clip]{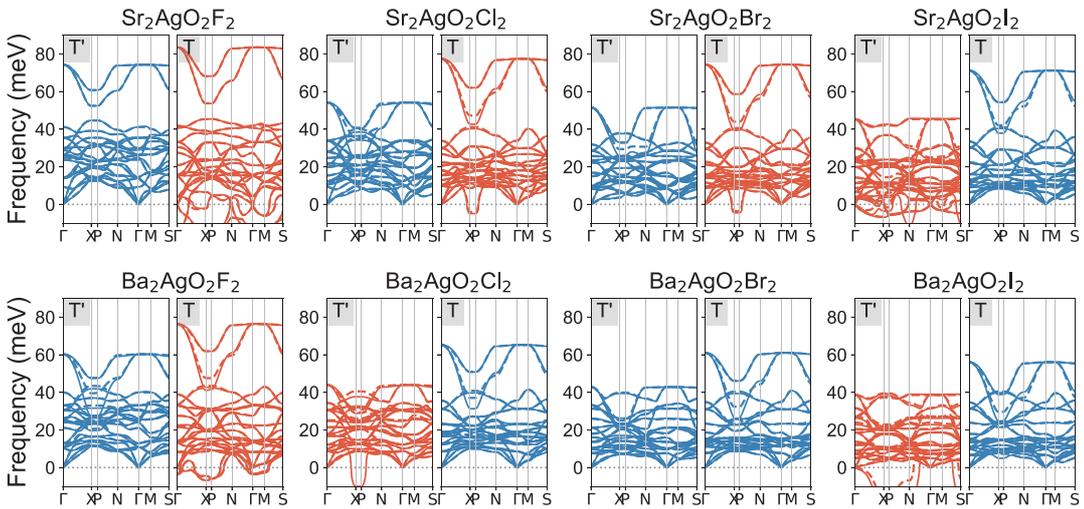}
	\caption{
			Phonon dispersion of the Ag-based oxides computed within harmonic approximation. The solid lines are results obtained using the 2$\times$2$\times$2 supercell, and the dashed lines represent the result obtained using a larger supercell containing 112 atoms. The imaginary frequencies are shown as negative values. Different colors are used for dynamically stable (blue) and unstable (red) systems.
	}
	\label{phonon_Ag}
\end{figure*}
	\begin{figure*}[htp]
		\centering
		\includegraphics[width=0.8\textwidth,clip]{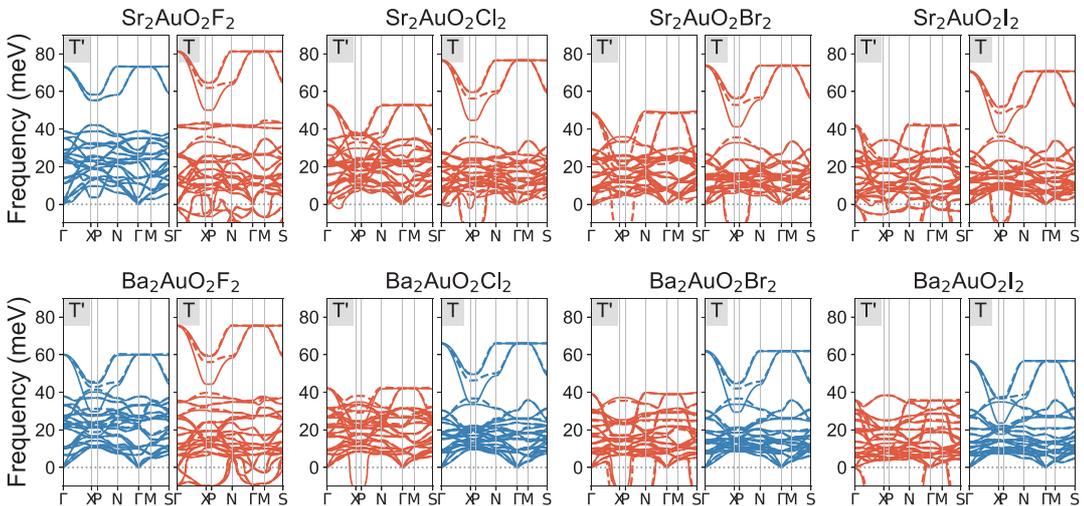}
		\caption{
				Phonon dispersion of the Au-based oxides computed within harmonic approximation. The meaning of the different line styles and colors are described in Fig.~\ref{phonon_Ag} caption.
		}
		\label{phonon_Au}
	\end{figure*}
Figure \ref{phonon_Ag} compares the calculated phonon dispersion curves of the T'- and T-type Ag-based oxides $A_2$AgO$_2X_2$. Among the 16 Ag-based oxides, nine oxides are predicted to be dynamically stable, which are shown with blue lines in Fig.~\ref{phonon_Ag}. 
For $X$=F, the T'-type structure is dynamically stable, whereas the T-type is dynamically unstable where imaginary phonon modes appear in the wide-range of the Brillouin zone (BZ). In the largest halogen ion case ($X$=I), the T-type is dynamically stable and the T'-type is unstable. This $X$-dependency of the dynamical stability is consistent with the energy difference between the T-type and T'-type phases mentioned above. For Sr$_2$AgO$_2$Cl$_2$ and Sr$_2$AgO$_2$Br$_2$, the T-type phase appears to be dynamically unstable even though it is energetically more stable than the T'-type structure. The soft phonon mode of the T-type phases, which appears only around the X and P points of the BZ, 
involves in-plane displacements of the corner-sharing oxygen atoms in the AgO$_2$ plane and induces the structural phase transition to a lower-symmetry orthorhombic phase, which resembles the unstable octahedral tilting mode in the tetragonal (T-type) La$_2$CuO$_4$~\cite{LCO_phonon}. Since the distortion of the tetragonal La$_2$CuO$_4$ to the orthorhombic phase can be suppressed by hole doping~\cite{LCO_hole_doping}, a similar doping approach is expected to improve the stability of the T-type Sr$_2$AgO$_2$Cl$_2$ and Sr$_2$AgO$_2$Br$_2$. For $A = $ Ba, the $X$-dependency of the dynamical stability is similar to the case of $A = $ Sr except for $X = $ Cl and Br where the T-type is predicted to be dynamically stable. The soft phonons observed in the T-type Sr$_2$AgO$_2X_2$ ($X = $ Cl, Br) are stabilized in the corresponding Ba$_2$AgO$_2X_2$, which can be attributed to the larger in-plane lattice constants of the Ba systems along with the negative Gr\"{u}neisen parameter of the soft mode.

For the Au-based oxide, the halogen-ion dependency of the lattice parameters, energy difference between the T'- and T-type structures, and the dynamical stability are the same as the Ag-based oxides, as shown in Tables ~\ref{table:structures_Tprime_Au} and ~\ref{table:structures_T_Au} in Appendix ~\ref{sec:structure_params_Au} and Fig.~\ref{phonon_Au}. However, only five structures are predicted to be dynamically stable.

\begin{table}[tb]
	\caption{Structural parameters of the studied T'-type $A_{2}$AuO$_{2}X_{2}$ (space group: $I4/mmm$) optimized with PBEsol functional. The $A$ atoms occupy the 4e Wyckoff site $(0,0,\pm z_A)$. The Au, O, and $X$ atoms are located at the 2a, 4c, and 4d sites, respectively.}
	\label{table:structures_Tprime_Au}
	\begin{ruledtabular}
		\begin{tabular}{lcccc}
			Composition & $a$ (\AA) & $c$ (\AA) & $c/a$ & $z$ \\
			\hline
			Sr$_{2}$AuO$_{2}$F$_2$ &  4.173 & 12.435 & 2.9802 & 0.3616 \\
			Sr$_{2}$AuO$_{2}$Cl$_2$ &  4.363 & 14.115 & 3.2352 & 0.3823 \\
			Sr$_{2}$AuO$_{2}$Br$_2$ &  4.445 & 14.760 & 3.3203 & 0.3920 \\
			Sr$_{2}$AuO$_{2}$I$_2$ &  4.587 & 15.691 & 3.4208 & 0.4043 \\
			Ba$_{2}$AuO$_{2}$F$_2$ &  4.276 & 13.707 & 3.2055 & 0.3639 \\
			Ba$_{2}$AuO$_{2}$Cl$_2$ &  4.456 & 15.327 & 3.4396 & 0.3843 \\
			Ba$_{2}$AuO$_{2}$Br$_2$ &  4.534 & 15.845 & 3.4944 & 0.3923 \\
			Ba$_{2}$AuO$_{2}$I$_2$ &  4.656 & 16.688 & 3.5845 & 0.4042 \\
		\end{tabular}
	\end{ruledtabular}
\end{table}

\begin{table}[h!]
	\caption{Structural parameters of the studied T-type $A_{2}$AuO$_{2}X_{2}$ (space group: $I4/mmm$) optimized with PBEsol functional. The $A$ and $X$ atoms occupy the 4e site $(0,0,\pm z)$, and Au atoms and O atoms are located at the 2a and 4c sites, respectively.}
	\label{table:structures_T_Au}
	\begin{ruledtabular}
		\begin{tabular}{lccccc}
			Composition & $a$ (\AA) & $c$ (\AA) & $c/a$ & $z_A$ & $z_B$ \\
			\hline
			Sr$_{2}$AuO$_{2}$F$_2$ &  4.126 & 13.558 & 3.286 & 0.3744 & 0.2022 \\
			Sr$_{2}$AuO$_{2}$Cl$_2$ &  4.175 & 15.474 & 3.707 & 0.3889 & 0.1918 \\
			Sr$_{2}$AuO$_{2}$Br$_2$ &  4.201 & 16.413 & 3.907 & 0.3960 & 0.1880 \\
			Sr$_{2}$AuO$_{2}$I$_2$ &  4.231 & 18.573 & 4.390 & 0.4102 & 0.1783 \\
			Ba$_{2}$AuO$_{2}$F$_2$ &  4.165 & 14.580 & 3.501 & 0.3716 & 0.1947 \\
			Ba$_{2}$AuO$_{2}$Cl$_2$ &  4.256 & 16.409 & 3.856 & 0.3870 & 0.1891 \\
			Ba$_{2}$AuO$_{2}$Br$_2$ &  4.296 & 17.177 & 3.999 & 0.3933 & 0.1878 \\
			Ba$_{2}$AuO$_{2}$I$_2$ &  4.350 & 18.663 & 4.290 & 0.4041 & 0.1840 \\
		\end{tabular}
	\end{ruledtabular}
\end{table}

\subsection{Selection of promising materials}

\begin{figure}[h!]
	\centering 
	\includegraphics[clip,width=0.4\textwidth ]{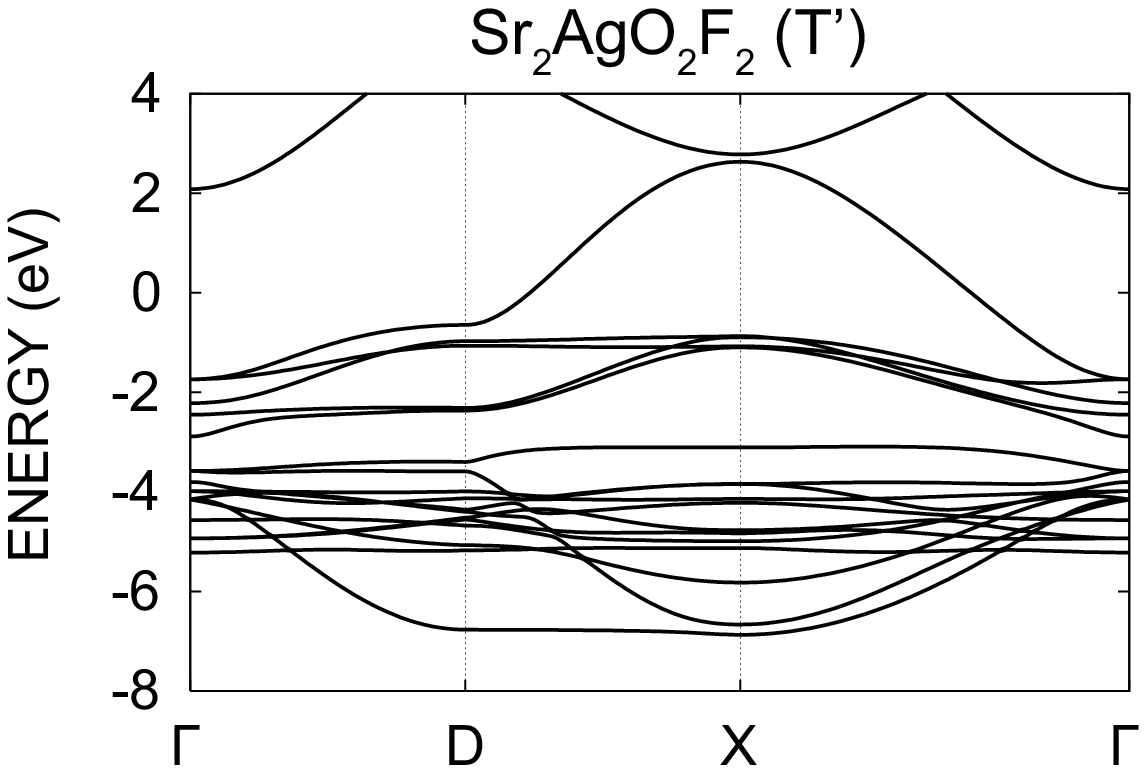} 
	\caption{(Color online) Electronic band structure of Sr$_2$AgO$_2$F$_2$ in the LDA.
		The zero energy corresponds to the Fermi level. 
	}
	\label{bndSAOF_LDA}
\end{figure} 
For all stable Ag- and Au-based oxides and the representative Cu-based oxides, we perform the band structure calculations in the DFT/LDA.
Figure~\ref{bndSAOF_LDA} shows a LDA band structure of one of them, for Sr$_2$AgO$_2$F$_2$.
As in the case of typical cuprates, a single band originating from the antibonding orbitals appears on the Fermi level in all the Ag- and Au-based oxides calculated in this study.
Therefore, as in the case of the copper oxides, it is possible to discuss the system in terms of a low-energy effective Hamiltonian consisting mainly of \tor{an antibonding orbital}.

\begin{table*}[th!]  
	\caption{
		On-site potentials and effective interactions for one-band Hamiltonian of Cu-, Ag-, and Au-based compounds (in eV).
		The one-body part is obtained from the fitting of the LDA band structure,
		while the effective interaction is the result of the cRPA from the LDA bands.
		$v$, $U$, and $V$ represent the bare Coulomb, the static values of the effective on-site Coulomb, and the effective off-site Coulomb, respectively (at $\omega=0$).
		$U/v$ is the ratio of the on-site bare and screened Coulomb interactions, and $U/t$ is the ratio of the nearest neighbor hopping and the screened Coulomb interaction.
	}
	\ 
	\label{para1_LDA} 
	\scalebox{0.92}[1.0]{
	\begin{tabular}{c|c|c|c|cc|cc|cc|cc|c|c} 
		\hline \hline \\ [-8pt]
		Cu & $t(1,0,0)$ &  $t(1,1,0)$  & $t(2,0,0)$ &  $v(0,0,0)$   & $U(0,0,0)$ & $v(1,0,0)$  & $V(1,0,0)$ & $v(1,1,0)$ & $V(1,1,0)$  & $v(2,0,0)$  & $V(2,0,0)$  &  $U/v$   &  $|U/t|$    \\ [+1pt]
		\hline \\ [-8pt] 
		Ca$_2$CuO$_2$Cl$_2$   &   -0.492 & 0.091 & -0.053  & 18.220 &5.063 &  4.170 &  1.183 & 2.768  &0.736 & 2.090  &0.534 &  0.278  & 10.29          \\
		\hline \\ [-8pt] 
		La$_2$CuO$_4$  &  -0.507 &0.084 & -0.051  & 18.354& 4.335& 4.241 & 0.665 & 2.825 &0.345 & 2.138  &0.231 &  0.236  & 8.55          \\
		\hline \\ [-8pt] 
		Sr$_2$CuO$_2$F$_2$  &   -0.489&0.105 & -0.049  & 18.449& 4.780&  4.088 &  0.962 &2.699 &0.542 &2.033  &0.361 &  0.259  & 9.78          \\
		\hline \\ [-8pt] 
		HgBa$_2$CuO$_4$  &  -0.502 &0.091& -0.063  &16.856& 3.957& 4.164 &  0.763 & 2.749 &0.407 &2.090  &0.270 &  0.235  & 7.88         \\
		\hline \\ [-8pt] 
		CaCuO$_2$  &  -0.512 &0.084 & -0.062  & 17.104& 3.999&  4.193 & 0.763 &2.758 &0.361 & 2.081  &0.204 &  0.234  & 7.81          \\  
		\hline \hline \\ [-8pt]
		Ag  & $t(1,0,0)$ &  $t(1,1,0)$  & $t(2,0,0)$ &  $v(0,0,0)$   & $U(0,0,0)$ & $v(1,0,0)$  & $V(1,0,0)$ & $v(1,1,0)$ & $V(1,1,0)$   & $v(2,0,0)$  & $V(2,0,0)$   &  $U/v$   &  $|U/t|$   \\ [+1pt]
		\hline \\ [-8pt] 
		Sr$_2$AgO$_2$F$_2$ (T')   &   -0.644 &0.111 &-0.120  & 11.046& 3.653&  3.974 &  1.239 & 2.607 &0.773 & 1.983  &0.590 &  0.331  & 5.67          \\
		\hline \\ [-8pt] 
		Sr$_2$AgO$_2$Cl$_2$ (T')  &   -0.545 &0.094 & -0.114  &10.310& 2.813&  3.802 &  0.730 & 2.487 &0.346& 1.900  &0.227 &  0.273  & 5.16         \\
		\hline \\ [-8pt] 
		Sr$_2$AgO$_2$I$_2$ (T)  &   -0.602 &0.103 & -0.106  & 10.988& 3.132& 3.904 &  0.790 & 2.566 &0.404 & 1.952  &0.278 &  0.285  & 5.20          \\
		\hline \\ [-8pt] 
		Ba$_2$AgO$_2$F$_2$ (T')  &  -0.596 &0.106 &-0.119  & 10.468 & 2.999&3.873 & 0.875 & 2.537 &0.473 & 1.936  &0.332 &  0.286  & 5.03         \\
		\hline \\ [-8pt] 
		Ba$_2$AgO$_2$Cl$_2$ (T)  &   -0.585 &0.099 & -0.110  & 10.675 & 3.323 &  3.871 &  1.078 & 2.542 &0.667 &1.939  &0.511 &  0.311  & 5.68          \\
		\hline \\ [-8pt]
		Ba$_2$AgO$_2$Br$_2$ (T)   &  -0.568 &0.100 & -0.109  & 10.512& 3.080 & 3.833 & 0.905 & 2.516 &0.511& 1.922  &0.369 & 0.293  & 5.42         \\
		\hline \\ [-8pt] 
		Ba$_2$AgO$_2$I$_2$ (T)  &  -0.547 &0.100 & -0.109 & 10.316 & 2.780& 3.788 &  0.699 & 2.485 &0.340 & 1.901  &0.225 &  0.269 & 5.08          \\  
		\hline \hline \\ [-8pt]
		Au  & $t(1,0,0)$ &  $t(1,1,0)$  & $t(2,0,0)$ &  $v(0,0,0)$   & $U(0,0,0)$ & $v(1,0,0)$  & $V(1,0,0)$ & $v(1,1,0)$ & $V(1,1,0)$   & $v(2,0,0)$  & $V(2,0,0)$   &  $U/v$   &  $|U/t|$   \\ [+1pt]
		\hline \\ [-8pt] 
		Ba$_2$AuO$_2$F$_2$ (T')  &  -0.674 &0.166 & -0.151  & 9.230 & 2.474 & 3.838 & 0.758 & 2.535 &0.433 & 1.987  &0.342  & 0.268  & 3.67          \\
		\hline \\ [-8pt] 
		Ba$_2$AuO$_2$Cl$_2$ (T)  &   -0.676 &0.162& -0.141 & 9.436 & 2.634 &  3.843 & 0.815 & 2.551 &0.467 & 1.987  &0.342 &  0.279  & 3.90         \\
		\hline \\ [-8pt]
		Ba$_2$AuO$_2$Br$_2$ (T)   &  -0.659 &0.160 & -0.141 & 9.319 & 2.454 &3.809 & 0.689 & 2.527 &0.366 & 1.975  &0.274  &0.263  & 3.72        \\
		\hline \\ [-8pt] 
		Ba$_2$AuO$_2$I$_2$ (T)  &   -0.633 &0.158 & -0.138  &9.221& 2.338 &  3.763 &  0.596 & 2.496 &0.282 & 1.954  &0.199  &  0.254  &3.69         \\  
		\hline
		\hline 
	\end{tabular} 
}
\end{table*} 
From the obtained band structure, following the standard MACE procedure, we construct one-band Hamiltonians of the antibonding orbital and three-band Hamiltonians of the $x^2-y^2$ and the $2p$ orbitals first on the LDA level to quickly pick up promising candidates to be examined later in depth.
We summarize thus obtained results of the most important parameters of the effective one-band Hamiltonian in Table~\ref{para1_LDA}.

See also Appendix \ref{BandLDA} for the corresponding Wannier function 
and for the fitting of one band for several other Ag-based oxides.
In the antibonding Wannier function, the $x^2-y^2$ orbital of Ag and the $2p$ orbitals of O are strongly hybridized.
Therefore, the bandwidth of the antibonding band is very large, 5-6 eV.

Here, we focus on the effective correlation strength $|U/t|$ listed up in Table~\ref{para1_LDA}.
Note that $U/t$ here is obtained from a simple fitting of the LDA band together with the cRPA, where $U$ is derived from the LDA band and $t$ is a simple fitting of the LDA band.
More precise estimates of $U$ and $t$ for several Ag-based oxides will be given later.
It is known~\cite{hirayama18,hirayama19} that the copper oxides have large $|U/t|$ of 8-10, and systems with smaller $|U/t|$ tend to have higher $T_c$.
HgBa$_2$CuO$_4$ and CaCuO$_2$ exhibiting relatively high $T_c$ indeed have $|U/t|\sim8$, which are near the lower boundary of the cuprates and it is desired to find lower $|U/t|$ to seek for higher $T_c$ by considering the possibility that the cuprates did not reach the optimized $U/t$.

For the Ag-based oxides we selected, however, $|U/t|$ is about 5-6.
Compared to the copper oxides, it provides us with an approach from relatively weaker correlation side.
The nearest-neighbor hopping of cuprate is about $-0.50$ eV, regardless of the compounds.
On the other hand, the silver oxides have larger hopping amplitudes than that of the copper oxides.
The hopping of silver oxide strongly depends on the materials, ranging from $-0.65$ eV to $-0.55$ eV, and has larger value roughly for smaller lattice constant (see Tables~\ref{table:structures_Tprime_Ag} and \ref{table:structures_T_Ag}).
Of course, larger $|t|$ is preferred to make the characteristic energy scale and the resultant possible superconducting gap and temperature scales higher.
We have also calculated the Au-based oxide for comparison.
The $|U/t|$ is about 3.5-4 and the system is not strongly correlated.

From these analyses, we pick up compounds that have either larger $|U/t|$ or larger $|t|$: Below we study in depth four Ag-based oxides, Sr$_2$AgO$_2$F$_2$, Sr$_2$AgO$_2$I$_2$, Ba$_2$AgO$_2$F$_2$, and Ba$_2$AgO$_2$Cl$_2$ for more detailed and accurate studies with heavier computational cost.
The choice of four compounds is also supported from the analyses in Appendix~\ref{BandLDA} for the three band Hamiltonian on the LDA level.

\subsection{Effective Hamiltonian derived at the cGW-SIC beyond the DFT/LDA level}

Now, we present the effective Hamiltonians obtained by using more accurate procedure of MACE beyond DFT/LDA
level.
For a while, we focus on the Ag-based oxides and compare with the cuprates.
We employ the standard procedure of the MACE~\cite{imadamiyake10,hirayama17,hirayama18,hirayama19}. In the cGW-SIC scheme, we first replace the LDA with the GWA, which \tor{enables} us more accurate calculation of the correlation effect of the bands outside of the low-energy Hamiltonian than the LDA. It also enables \tor{removing} the double counting of the correlation effect in the effective Hamiltonian, which is unavoidable when we rely on the LDA. 
Figure~\ref{bndAgGW} in Appendix~\ref{BandGWA} shows the band structures obtained in the GWA which is used in the next step of the cGW-SIC.
The effective Hamiltonian at the GW level as the intermediate step is described in Appendix~\ref{BandGWA}.

The final and most accurate effective Hamiltonian is given in the cGW-SIC level by using the GW band structure obtained above as the starting point and we present one-band Hamiltonian for the antibonding band and the three-band Hamiltonian as well by using the two bases: One, basis of both $d_{x^2-y^2}$ and two $2p$ orbitals ($dpp$ basis) and the other, the basis in the antibonding
and two bonding orbitals (abb basis).

\subsubsection{Derivation of single-band Hamiltonian}
\begin{figure*}[htp]
	\includegraphics[clip,width=0.7\textwidth ]{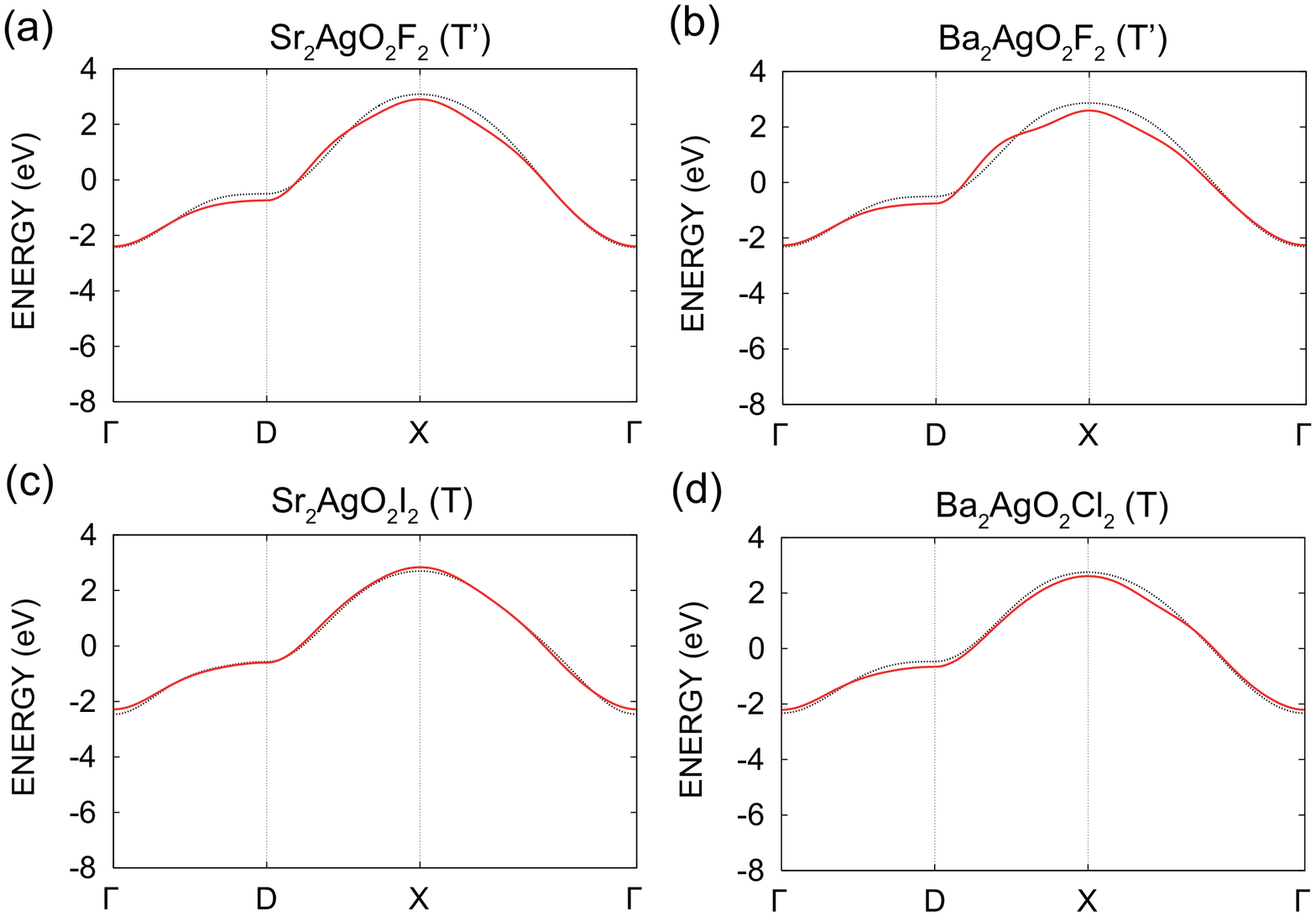} 
	\caption{(Color online) Electronic band structure of one-band Hamiltonian in the cGW 
		originating from the anti-bonding Wannier orbitals.
		The zero energy corresponds to the Fermi level. 
		For comparison, the fitting band structure for the GWA is also given (black dotted line).
	}
	\label{bndAgcGWwan1}
\end{figure*}
First, as in the LDA calculations, we summarize the parameters of the \tor{one-band} Hamiltonian at the cGW level, by using the hopping calculated from the maximally localized Wannier orbitals constructed from the GW band and the effective interaction obtained by the cRPA, in Table~\ref{paraAg1_cGW} for the four Ag compounds~\cite{LR}.
\mi{We summarize the details of the \tor{one-band} parameters in Supplemental Materials~\cite{smdata}.} 
Figure~\ref{bndAgcGWwan1} shows the corresponding band structure at the cGW level.
The $|U/t|$ values are larger than that in the LDA due to the self-energy effect, with the largest value being about 6.69 in Sr$_2$AgO$_2$F$_2$,
which is larger than the simplified derivation 5.67 in Table~\ref{para1_LDA}.
The nearest-neighbor hopping of Sr$_2$AgO$_2$F$_2$ is also $-0.708$ eV, which is larger than that in the LDA fitting  $-0.644$ eV, and the energy scale is increased.

\begin{table*}[htp] 
	\caption{
		Transfer integral and effective interaction in one-band Hamiltonian for Sr$_2$AgO$_2$F$_2$, Sr$_2$AgO$_2$I$_2$, Ba$_2$AgO$_2$F$_2$, and Ba$_2$AgO$_2$Cl$_2$ (in eV).
		Transfer integrals and effective interactions are obtained by using the cGW and the cRPA applied to the GW Green's function, respectively.
		$v$, $U$, and $V$ represent the bare Coulomb, the static values of the effective on-site Coulomb, and the effective off-site Coulomb, respectively (at $\omega=0$).
The parameters for further neighbor transfer integrals and interactions by the cGW 
		are given \mi{in Table~\ref{SrAgOF_full}  for  Sr$_2$AgO$_2$F$_2$ and for other compounds, in the Supplemental Material~\cite{smdata}.}
	}
	\ 
	\label{paraAg1_cGW} 
	\scalebox{0.92}[1.0]{
	\begin{tabular}{c|c|c|c|cc|cc|cc|cc|c|c} 
		\hline \hline \\ [-8pt]
		cGW  & $t(1,0,0)$ &  $t(1,1,0)$  & $t(2,0,0)$ &  $v(0,0,0)$   & $U(0,0,0)$ & $v(1,0,0)$  & $V(1,0,0)$ & $v(1,1,0)$ & $V(1,1,0)$    &  $v(2,0,0)$ & $V(2,0,0)$    & $U/v$   &  $|U/t|$   \\ [+1pt]
		\hline \\ [-8pt] 
		Sr$_2$AgO$_2$F$_2$ (T') &-0.658 &  0.092 &  -0.101 &  11.964  &4.401 &  3.982  & 1.464 & 2.603  & 0.966 & 1.961  & 0.771 & 0.368 & 6.69          \\
		\hline \\ [-8pt] 
		Ba$_2$AgO$_2$F$_2$ (T')  &  -0.609 &0.067 &-0.093 & 11.485 & 4.010 &3.894 & 1.282 & 2.533 &0.811 & 1.911  & 0.629 &  0.349  & 6.58         \\
		\hline \\ [-8pt] 
		Sr$_2$AgO$_2$I$_2$ (T)  &   -0.626 &0.101 & -0.093  & 11.063& 3.337& 3.907 &  0.938 & 2.564 &0.535 & 1.951  & 0.393 & 0.302  & 5.33          \\
		\hline \\ [-8pt] 
		Ba$_2$AgO$_2$Cl$_2$ (T)  &   -0.589 &0.093  &-0.099  & 11.434 & 3.728 &  3.881 &  1.115 & 2.536 &0.694 & 1.914  & 0.535 &  0.326  & 6.33          \\
		\hline
		\hline 
	\end{tabular} 
}
\end{table*} 

\subsubsection{Derivation of three-band $dpp$ Hamiltonian}

\begin{table*}[htp]
	\caption{
		Transfer integrals and effective interactions for three-band $dpp$ Hamiltonian of Sr$_2$AgO$_2$F$_2$ (T') (in eV).
		We show the transfer integral in the cGW-SIC, while the effective interaction is the result of the cRPA.
		$v$ and $J_{v}$ represent the bare Coulomb and exchange interactions, respectively.
		$U$, $V$, and $J$ represent the static values of the on-site effective Coulomb, the off-site effective Coulomb, and exchange interactions, respectively (at $\omega=0$).
		The occupation number in the GWA is also given in the bottom column ``occu.(GWA)'' in this Table.
		The parameters for further neighbor transfer integrals and interactions by the cGW-SIC are given in the Supplemental Material~\cite{smdata}.
	}
	\
	\label{paraAg3cGW-SIC_SAOF} 
	\begin{tabular}{c|ccc|ccc|ccc|ccc} 
		\hline \hline \\ [-8pt]
		cGW-SIC   &       &  $t(0,0,0)$  &       &     & $t(1,0,0)$ &    &       & $t(1,1,0)$ &      &     &  $(2,0,0)$ &     \\ [+1pt]
		\hline \\ [-8pt]
		&  $x^2-y^2$ &  $p_1$ &  $p_2$ & $x^2-y^2$ &  $p_1$ &  $p_2$ &  $x^2-y^2$ &  $p_1$ &  $p_2$ & $x^2-y^2$ &  $p_1$ &  $p_2$ \\ 
		\hline \\ [-8pt] 
		$x^2-y^2$  & -3.091& -1.562 & 1.562   &  -0.003 &-0.059 &-0.070 &  0.028 &-0.015 & 0.015 &  -0.023  &0.008& -0.003 \\ 
		$p_1$   & -1.562 &-3.702 &-0.688  &  1.562 & 0.288 & 0.688  & -0.070 & 0.061&  0.011 & 0.059 &-0.019 & 0.011 \\
		$p_2$    &  1.562 &-0.688 &-3.702  &   -0.070& -0.011& -0.060 &  0.070 & 0.011&  0.061 & -0.003 & 0.002 & 0.000    \\
		\hline \hline \\ [-8pt]  
		&       &  $v(0,0,0)$  &       &     & $U(0,0,0)$ &    &       & $J_{v}(0,0,0)$ &      &     &  $J(0,0,0)$ &     \\ [+1pt]
		\hline \\ [-8pt]
		&  $x^2-y^2$ &  $p_1$ &  $p_2$ & $x^2-y^2$ &  $p_1$ &  $p_2$ &  $x^2-y^2$ &  $p_1$ &  $p_2$ & $x^2-y^2$ &  $p_1$ &  $p_2$ \\ 
		\hline \\ [-8pt] 
		$x^2-y^2$ & 19.455& 7.519 &7.519   & 8.493 &2.865& 2.865 &            &  0.084 &   0.084  &           &   0.064 &  0.064  \\ 
		$p_1$         &  7.519& 17.012 &5.030  & 2.865& 6.213& 1.891  &  0.084  &            &  0.043  & 0.064  &            &  0.020 \\
		$p_2$        &   7.519& 5.030& 17.012  &2.865& 1.891 &6.213 & 0.084   & 0.043  &            &  0.064  & 0.020  &         \\
		\hline \hline \\ [-8pt]  
		&       &  $v(1,0,0)$ &    &     & $V(1,0,0)$ &    &       & $v(1,1,0)$  &      &     &  $V(1,1,0)$ &     \\ [+1pt]
		\hline \\ [-8pt] 
		&  $x^2-y^2$ &  $p_1$ &  $p_2$ & $x^2-y^2$ &  $p_1$ &  $p_2$ &  $x^2-y^2$ &  $p_1$ &  $p_2$ & $x^2-y^2$ &  $p_1$ &  $p_2$ \\ 
		\hline \\ [-8pt] 
		$x^2-y^2$ &   3.585 &  7.519 &  3.149  & 1.466 &  2.865 &  1.214   &  2.545 &  3.152&   3.152  &  0.991  & 1.216 &  1.216  \\
		$p_1$         &  2.431  & 3.681 &  2.283  & 1.015  & 1.557 &  0.911  &   2.050 &  2.536 &  2.287  & 0.814   &0.994&   0.913 \\
		$p_2$        &   3.152  & 5.032 &  3.411   &  1.216  & 1.893  & 1.264  & 2.050 &  2.287&   2.536  & 0.814 &  0.913 &  0.994 \\
		\hline \\ [-8pt]
		occ.(GWA)      &  $x^2-y^2$ &  $p_1$ &  $p_2$  \\ 
		\hline \\ [-8pt] 
		& 1.544 & 1.728  & 1.728   \\
		\hline
		\hline 
	\end{tabular}
\end{table*} 
\begin{table*}[htp]
	\caption{
		Transfer integrals and effective interactions for three-band abb Hamiltonian of Sr$_2$AgO$_2$F$_2$ (T') in the abb gauge (in eV).
		We show the transfer integral in the GW and the cGW-SIC, while the effective interaction is the result of the cRPA.
		The Wannier center of one antibonding and two bonding orbitals are the same as those of the $d$ and $p$ orbitals.
		$v$ and $J_{v}$ represent the bare Coulomb and exchange interactions, respectively.
		$U$, $V$, and $J$ represent the static values of the on-site effective Coulomb, the off-site effective Coulomb, and exchange interactions, respectively (at $\omega=0$).
		The occupation number in the GWA is also given in the bottom column ``occu.(GWA)'' in this Table.
	}
	\
	\label{paraAg3abb} 
	\begin{tabular}{c|ccc|ccc|ccc|ccc} 
		\hline \hline \\ [-8pt]
		  GW   &       &  $t(0,0,0)$  &       &     & $t(1,0,0)$ &    &       & $t(1,1,0)$ &      &     &  $t(2,0,0)$ &     \\ [+1pt]
		\hline \\ [-8pt]
		&  $a$ &  $b_1$ &  $b_2$ & $a$ &  $b_1$ &  $b_2$ &  $a$ &  $b_1$ &  $b_2$ & $a$ &  $b_1$ &  $b_2$ \\ 
		\hline \\ [-8pt] 
		$a$  & 0.102&0.000 &0.000   &  -0.615 &0.000 &0.000 &  0.080 &0.000 & 0.000 &  -0.130  &0.000& 0.000 \\ 
		$b_1$   & 0.000 &-5.140 &0.135 &  0.000 & 0.732 & -0.136  & 0.000 & 0.006&  -0.051 & 0.000 &0.148 & -0.051 \\
		$b_2$    & 0.000 &0.135 &-5.140  &   0.000& 0.051& -0.009 &  0.000 & -0.051& 0.006 & 0.000 & 0.010 & -0.020    \\
		\hline \hline \\ [-8pt]
		cGW-SIC   &       &  $t(0,0,0)$  &       &     & $t(1,0,0)$ &    &       & $t(1,1,0)$ &      &     &  $t(2,0,0)$ &     \\ [+1pt]
		\hline \\ [-8pt]
		&  $a$ &  $b_1$ &  $b_2$ & $a$ &  $b_1$ &  $b_2$ &  $a$ &  $b_1$ &  $b_2$ & $a$ &  $b_1$ &  $b_2$ \\ 
		\hline \\ [-8pt] 
		$a$  & 0.062& -0.055 & -0.055   & -0.631 &-0.056 &-0.024 & 0.141 &0.002 & -0.002 &  -0.192  &-0.010 &  0.006 \\ 
		$b_1$   & -0.055 &-5.279 &0.107  & 0.055& 0.845 & -0.107  & -0.024 &0.005& -0.054 & 0.056 &0.164 &  -0.054 \\
		$b_2$    &   -0.055 &0.107 &-5.279  &-0.024& 0.054& 0.011 & 0.024 & -0.054& 0.005 &0.006 & 0.010 & -0.013    \\
		\hline \hline \\ [-8pt]  
		&       &  $v(0,0,0)$  &       &     & $U(0,0,0)$ &    &       & $J_{v}(0,0,0)$ &      &     &  $J(0,0,0)$ &     \\ [+1pt]
		\hline \\ [-8pt]
		&  $a$ &  $b_1$ &  $b_2$ & $a$ &  $b_1$ &  $b_2$ &  $a$ &  $b_1$ &  $b_2$ & $a$ &  $b_1$ &  $b_2$ \\ 
		\hline \\ [-8pt] 
		$a$   & 10.045& 7.571 &7.571   & 4.088 &2.928& 2.928 &            &  1.738 &  1.738  &           &   0.703 &  0.703  \\ 
		$b_1$ &  7.571& 12.902 &5.095  & 2.928&4.924&1.944  &  1.738  &            &   0.153  & 0.703  &            &  0.072 \\
		$b_2$ &   7.571& 5.095& 12.902  &2.928& 1.944 &4.924 & 1.738   &  0.153  &            &  0.703  & 0.072  &         \\
		\hline \hline \\ [-8pt]  
		&       &  $v(1,0,0)$ &    &     & $V(1,0,0)$ &    &       & $v(1,1,0)$  &      &     &  $V(1,1,0)$ &     \\ [+1pt]
		\hline \\ [-8pt] 
		&  $a$ &  $b_1$ &  $b_2$ & $a$ &  $b_1$ &  $b_2$ &  $a$ &  $b_1$ &  $b_2$ & $a$ &  $b_1$ &  $b_2$ \\  
		\hline \\ [-8pt] 
		$a$   &   3.976 &  7.571 &  3.193  & 1.540 &   2.928 & 1.231   &  2.604 & 3.196&   3.196  &   1.013  & 1.233 &  1.233  \\
		$b_1$ &   2.730  &  4.422 &  2.347  &1.115  & 1.821 &  0.937  &   2.105 &  2.550 &   2.351  & 0.834   &0.999&   0.939 \\
		$b_2$ &   3.196  & 5.097 &  3.259   & 1.233  &1.946  & 1.228  & 2.105 &   2.351&   2.550  & 0.834 &  0.939 &  0.999 \\
		\hline \\ [-8pt]
		occ.(GWA)      &  $a$ &  $b_1$ &  $b_2$  \\ 
		\hline \\ [-8pt] 
		& 1.000 & 2.000  & 2.000   \\
		\hline
		\hline 
	\end{tabular}
\end{table*} 

We also present the \tor{three-band} Hamiltonian 
%
 using the cGW-SIC method.
Unlike the one-body terms calculated by simple LDA, 
there are less drawbacks such as double counting of low-energy degrees of freedom.
We summarize in Table~\ref{paraAg3cGW-SIC_SAOF}  detailed \tor{three-band} parameters of the effective Hamiltonian at the cGW-SIC level based on the Wannier basis of Cu $d$ and O $p$ basis for Sr$_2$AgO$_2$F$_2$~\cite{LR}. 
\mi{We ignore the level renormalization taken into account before~\cite{hirayama19} because its effect is small. 
We summarize the details of the \tor{three-band} parameters in Supplemental Materials~\cite{smdata}.} 
In Appendix~\ref{dpp_cGW-SIC}, we show comparison of the band dispersions and effective Hamiltonians of four Ag-based compounds.


\subsubsection{Derivation of three-band abb Hamiltonian}
The three band Hamiltonian ($dpp$ Hamiltonian) represented in the $d$ and $p$ Wannier orbitals above can alternatively be represented by an antibonding orbital 
and two bonding orbitals.
When calculating with the VMC method, it is better to transform the basis to such a basis, since the Gutzwiller factor introduces only diagonal terms, where the most important local interaction at the antibonding band is straightforwardly and efficiently taken into account.
In this paper, we refer to this effective Hamiltonian as the abb Hamiltonian.
In Table.~\ref{paraAg3abb}, we show the GW and cGW-SIC parameters for the Sr$_2$AgO$_2$F$_2$ in the abb Hamiltonian. \mi{We also
summarize the details of the \tor{three-band} parameters in the
abb-gauge in Supplemental Materials~\cite{smdata}.}
For the $dpp$ band in the GWA, we apply the method of the maximally localized Wannier function to each of the anti-bonding band and the other two bands, and get the Wannier function in the abb Hamiltonian.
For hopping that reproduces the GW band, the anti-bonding orbital and the two bonding orbitals are orthogonal.
In the cGW-SIC, however, the anti-bonding orbital and the two bonding orbitals are not strictly orthogonal, but the hoppings between them are small  ($\sim$ 55 meV).
We also note that the two bonding orbitals are not orthogonal even at the GW level, because the Wannier orbitals are constructed to satisfy the maximally localized nature, which means that the two bonding orbitals here is represented by the linear combination of the orthognal bonding and nonbonding states.
Since the $d$ and $p$ orbitals form a strong covalent bond, the energy difference between the anti-bonding orbital and the two bonding orbitals is very large ($\sim$ 5 eV) due to the hybridization gap.
Therefore, the screening effect from the bonding and nonbonding orbitals for the antibonding electrons, which have small hopping and large energy difference with the bonding/nonbonding states, would be small.
The values of hopping and interaction of the antibonding orbital in the abb Hamiltonian are similar to the one-band Hamiltonian except for some reduction of the interaction and the modification of the longer ranged part of the hopping understood as the screening and the self-energy effects from the bonding/nonbonding electrons.
The reason why we do not employ the basis that diagonalizes the cGW-SIC band instead of the GW band is that the GW band should express rather better overall electronic structure after considering the correlation of the three bands.
Because of the large gap between the bonding and antibonding bands, the two bonding orbitals are nearly fully filled and only the antibonding band becomes partially filled, which makes the abb representation superior to the $dpp$ basis. This also means the single-band Hamiltonian is a good approximation and we solve it by the VMC below.

\subsection{Ground state of low-energy Hamiltonians}
	\begin{figure}[htp]
		\centering 
 		\includegraphics[width=0.4\textwidth ]{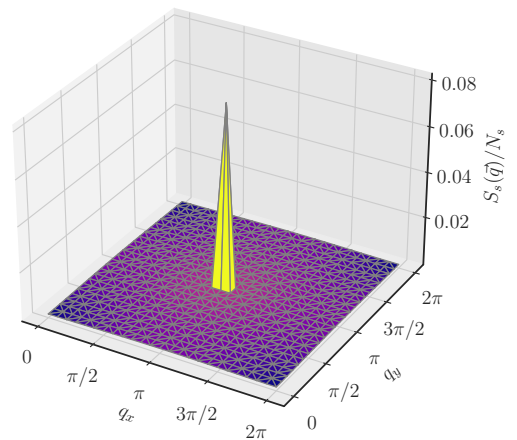} 
		\caption{(Color online) 
			Momentum dependence of spin structure factor $S_s(\bm{q})$ of non-doped Hamiltonian for Sr$_2$AgO$_2$F$_2$ calculated for \mi{a $24\times 24$ lattice}.
		}   
		\label{SQ}
	\end{figure} 
	\begin{figure}[htp]
		\centering 
 		\includegraphics[width=0.4\textwidth ]{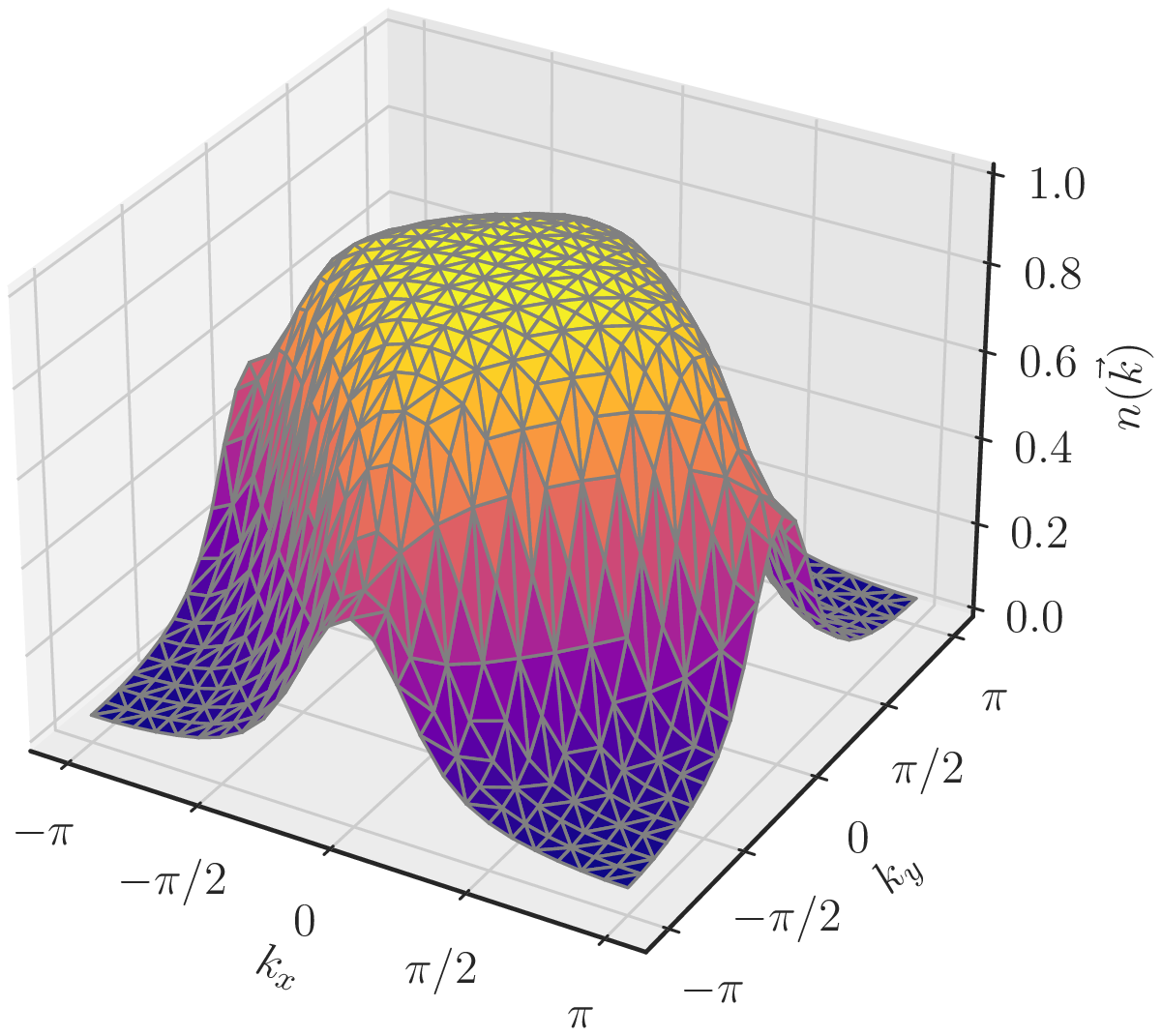} 
		\caption{(Color online) 
			Momentum distribution $n(\bm{k})$ of non-doped Hamiltonian for Sr$_2$AgO$_2$F$_2$ calculated for \mi{a $24\times 24$ lattice}.
		}   
		\label{nk}
	\end{figure} 	
\mi{Here we show a preliminary result of the ground state for the {\it ab initio} single-band effective Hamiltonian of Sr$_{2}$AgO$_2$F$_2$ on the cGW-SIC level obtained by VMC calculations to \mib{gain insight into Ag-based compounds whether} our proposed materials show} promising strong correlation effects. 
In the VMC calculation, we take into account the transfer integral up to the distance $(m,n)$ on the two-dimensional square lattice with $m,n \le 3$. For the interaction, we also take into account in the same range but for $(3,n)$, we fit from the result of $(2,n)$ in the form $a/\sqrt{m^2+n^2}$ with a constant $a$ because the $1/r$ fitting is known to be satisfactory already at $(2,n)$ (see Appendix B of Ref.~\onlinecite{ohgoe20}).  
In the solution obtained by the VMC calculation, two states are severely competing in the stoichiometric mother material, Sr$_{2}$AgO$_2$F$_2$; the G-type antiferromagnetic state with the normal staggered Neel order,
and correlated paramagnetic metal. 
The true ground state is the antiferromagnetic state with the energy \mts{ $E_{\rm AF}\sim $15.238 eV (per Ag site) in comparison to $E_{\rm metal}\sim$15.241 eV} \mi{for the paramagnetic metal 
after the 
variance extrapolation for $24\times 24$ lattice. The ground state is indeed characterized by the Bragg peak at $\bm{Q}=(\pi,\pi)$ of the spin structure factor defined in Eq.(\ref{Sq}) as is shown in Fig.~\ref{SQ}.
The ordered moment $m$ in Eq.~(\ref{m})  has a nonzero value $\sim 0.56$ slightly smaller than the case of La$_2$CuO$_4$\cite{hirayama19}. The momentum distribution defined in Eq.~(\ref{n_k}) for the} \mts{antiferromagnetic} \mi{ground state is given in Fig.~\ref{nk} for the $24 \times24$ lattice, clearly indicating the insulating nature with the absence of the  Fermi surface.
Therefore,} \mts{the} \mi{mother material is characterized by the antiferromagnetic Mott insulator, which is similar to the cuprates. However, the metallic state is severely competing with the antiferromagnetic insulator in the present case due to relatively weak} \mts{correlations.}

\section{Concluding Remark}
\label{Concluding}
\mi{We have demonstrated that the present family of silver-based compounds potentially shows strong correlation effects  similar to the cuprates. 
In the example of Sr$_{2}$AgO$_2$F$_2$, the non-doped mother compound shows  the antiferromagnetic Mott insulating ground state with the ordered moment comparable to La$_2$CuO$_4$. 
It offers promising candidates of the superconductivity when carriers are doped. }

It should be noted that Sr$_{2}$AgO$_2$F$_2$ has the largest off-site Coulomb interaction in the present family, while Sr$_{2}$AgO$_2$I$_2$ and Ba$_{2}$AgO$_2$Cl$_2$ have substantially weaker off-site interaction, which may enhance the superconductivity~\cite{ohgoe20} though $|U/t|$ is slightly smaller. In this sense Ag compounds open possibility of controlling parameters that affect the competition between superconductivity and other states, \mib{which contributes to the comprehensive understanding of the correlation induced superconductivity.} More thorough studies on the possible superconductivity will be reported elsewhere.  It is desired to investigate this family of materials experimentally after confirming that these materials can really be synthesized in accordance with the present prediction.

\begin{acknowledgments}
The authors acknowledge Youhei Yamaji, Yusuke Nomura, and Kota Ido for useful discussions.
This work was supported in part by KAKENHI Grant No. 16H06345 from JSPS. 
This research was also supported by MEXT as ``program for Promoting Researches on the Supercomputer Fugaku"(Basic Science for Emergence and Functionality in Quantum Matter - Innovative Strongly Correlated Electron Science by Integration of Fugaku and Frontier Experiments -, JPMXP1020200104). 
We thank the Supercomputer Center, the Institute for Solid State Physics, The University of Tokyo for the use of the facilities.
We also thank the computational resources of supercomputer Fugaku provided by the RIKEN Center for Computational Science (Project ID: hp210163, hp220166) and Oakbridge-CX in the Information Technology Center, The University of Tokyo. 
M.H. was supported by PRESTO, JST (JPMJPR21Q6). \\
\end{acknowledgments}

%
%
\appendix
\label{Appendix}

\section{Computational Conditions}
\label{CompCond}
\subsection{Conditions for phonon calculation}

The cell parameters and internal coordinates of the Ag- and Au-based oxides are fully relaxed using the \textit{Vienna Ab initio Simulation Package} (VASP)~\cite{vasp}, which implements the projector augmented wave (PAW) method~\cite{paw,paw2}. The VASP recommended PAW potentials for calculations based on the DFT are used together with the kinetic-energy cutoff of 500 eV for the plane wave expansion. The $k$-point mesh for the Brillouin zone integration is generated automatically so that the mesh density becomes as high as \tor{450 $\mathrm{\AA}^{3}$}. For the exchange-correlation potential, we employ the PBEsol functional within the generalized gradient approximation (GGA)~\cite{pbesol}, which gives better prediction accuracy of the structural parameters than GGA-PBE~\cite{pbe} and the LDA. 
The dynamical stability of each material is then accessed from the presence or absence of imaginary phonons in the phonon dispersion obtained within the harmonic approximation. To this end, we calculate phonon dispersion curves using the 2$\times$2$\times$2 supercells (containing 56 atoms) of the fully optimized structures using the finite displacement method, as implemented in the ALAMODE software~\cite{alamode}. If an unstable phonon, whose squared frequency is negative ($\omega_{\bm{q}\nu}^{2}<0$), is found on the commensurate 2$\times$2$\times$2 $\bm{q}$ points, the material is identified to be dynamically unstable; otherwise the system is dynamically stable. We further confirm the dynamical stability by using a larger supercell containing 112 atoms, which gives the quantitatively similar results as \tor{shown in Figs. 6 and 7}.

\subsection{Conditions for DFT and GW}

We calculate the electronic structure of the Cu-based compound using the experimental lattice parameters.
We employ the experimental results reported by Ref.~\onlinecite{Putilin} for HgBa$_2$CuO$_4$,
those reported by Ref.~\onlinecite{Jorgensen} for La$_2$CuO$_4$,
those reported by Ref.~\onlinecite{Hiroi} for Ca$_2$CuO$_2$Cl$_2$,
those reported by Ref.~\onlinecite{Kissick} for Sr$_2$CuO$_2$F$_2$,
and those reported by Ref.~\onlinecite{Karpinski} for CaCuO$_2$.
We calculate the electronic structure of the Ag-based and Au-based compounds using the lattice constants obtained from the structural optimization.

Computational conditions for the DFT/LDA and GW are as follows.
The band structure calculation is based on the full-potential
linear muffin-tin orbital (LMTO) implementation~\cite{methfessel}.
The exchange correlation functional is obtained by LDA of the Ceperley-Alder type.~\cite{ceperley}
We neglect the spin-polarization.
The self-consistent LDA calculation is done for the 12 $\times$ 12  $\times$ 12 $k$-mesh.
The muffintin (MT) radii are as follows:
$R^{\text{MT}}_{\text{Ca(Ca2CuO2Cl2)}}=$ 2.8 bohr,
$R^{\text{MT}}_{\text{Cu(Ca2CuO2Cl2)}}=$ 2.1 bohr,
$R^{\text{MT}}_{\text{O(Ca2CuO2Cl2)}}=$ 1.55 bohr,
$R^{\text{MT}}_{\text{Cl(Ca2CuO2Cl2)}}=$ 2.8 bohr,
$R^{\text{MT}}_{\text{La(La2CuO4)}}=$ 2.88 bohr,
$R^{\text{MT}}_{\text{Cu(La2CuO4)}}=$ 2.09 bohr,
$R^{\text{MT}}_{\text{O1(La2CuO4)}}=$ 1.40 bohr (in CuO$_2$ plane), $R^{\text{MT}}_{\text{O2(La2CuO4)}}=$ 1.60 bohr (others).
$R^{\text{MT}}_{\text{Sr(Sr2CuO2F2)}}=$ 3.1 bohr,
$R^{\text{MT}}_{\text{Cu(Sr2CuO2F2)}}=$ 2.16 bohr,
$R^{\text{MT}}_{\text{O(Sr2CuO2F2)}}=$ 1.55 bohr,
$R^{\text{MT}}_{\text{F(Sr2CuO2F2)}}=$ 1.55 bohr,
$R^{\text{MT}}_{\text{Hg(HgBa2CuO4)}}=$ 2.6 bohr,
$R^{\text{MT}}_{\text{Ba(HgBa2CuO4)}}=$ 3.6 bohr, 
$R^{\text{MT}}_{\text{Cu(HgBa2CuO4)}}=$ 2.15 bohr,
$R^{\text{MT}}_{\text{O1(HgBa2CuO4)}}=$ 1.50 bohr (in CuO$_2$ plane), 
$R^{\text{MT}}_{\text{O2(HgBa2CuO4)}}=$ 1.10 bohr (others),
$R^{\text{MT}}_{\text{Ca(CaCuO2)}}=$ 3.0 bohr,
$R^{\text{MT}}_{\text{Cu(CaCuO2)}}=$ 2.0 bohr,
$R^{\text{MT}}_{\text{O(CaCuO2)}}=$ 1.5 bohr,
$R^{\text{MT}}_{\text{Sr(Sr2AgO2F2)}}=$ 3.1 bohr,
$R^{\text{MT}}_{\text{Ag(Sr2AgO2X2)}}=$ 2.36 bohr,
$R^{\text{MT}}_{\text{O(A2AgO2X2)}}=$ 1.55 bohr,
$R^{\text{MT}}_{\text{F(A2AgO2F2)}}=$ 1.55 bohr,
$R^{\text{MT}}_{\text{Sr(Sr2AgO2Cl2)}}=$ 3.0 bohr,
$R^{\text{MT}}_{\text{Cl(Sr2AgO2Cl2)}}=$ 2.45 bohr,
$R^{\text{MT}}_{\text{Sr(Sr2AgO2Br2)}}=$ 2.95 bohr,
$R^{\text{MT}}_{\text{Br(A2AgO2Br2)}}=$ 2.4 bohr,
$R^{\text{MT}}_{\text{Sr(Sr2AgO2I2)}}=$ 3.0 bohr,
$R^{\text{MT}}_{\text{I(A2AgO2I2)}}=$ 2.6 bohr,
$R^{\text{MT}}_{\text{Ba(Ba2AgO2F2)}}=$ 3.1 bohr,
$R^{\text{MT}}_{\text{Ag(Ba2AgO2F2)}}=$ 2.47 bohr,
$R^{\text{MT}}_{\text{Ba(Ba2AgO2Cl2)}}=$ 3.0 bohr,
$R^{\text{MT}}_{\text{Ag(Ba2AgO2Cl2)}}=$ 2.4 bohr,
$R^{\text{MT}}_{\text{Cl(Ba2AgO2Cl2)}}=$ 3.1 bohr,
$R^{\text{MT}}_{\text{Ba(Ba2AgO2Br2 (T'))}}=$ 3.1 bohr,
$R^{\text{MT}}_{\text{Ag(Ba2AgO2Br2 (T'))}}=$ 2.7 bohr,
$R^{\text{MT}}_{\text{Ba(Ba2AgO2Br2 (T))}}=$ 3.0 bohr,
$R^{\text{MT}}_{\text{Ag(Ba2AgO2Br2 (T))}}=$ 2.5 bohr,
$R^{\text{MT}}_{\text{Ba(Ba2AgO2I2)}}=$ 3.0 bohr,
$R^{\text{MT}}_{\text{Ca(Ca2AgO2F2)}}=$ 2.8 bohr,
$R^{\text{MT}}_{\text{Ag(Ca2AgO2F2)}}=$ 2.26 bohr,
where A=Ca, Sr, and Ba, and X=F, Cl, Br, and I.
The angular momentum of the atomic orbitals is taken into account up to
$l=4$ for all the atoms.

The cRPA and GW calculations use a mixed basis consisting of products of two atomic orbitals and interstitial plane waves~\cite{schilfgaarde06}.
Because the basis of the calculation is not the plane wave but the muffin-tin orbitals, the interaction near the nuclei is well described,
which assures the stability of the calculation and the accuracy of the bare Coulomb interaction $v$ and the two-body term of the effective Hamiltonians.
In the cRPA and GW calculations, the 6 $\times$ 6 $\times$ 3 $k$-mesh is employed for Ca$_2$CuO$_2$Cl$_2$, La$_2$CuO$_4$, Sr$_2$CuO$_2$F$_2$, and Ag- and Au-based oxides, while 6 $\times$ 6 $\times$ 4 $k$-mesh is employed for HgBa$_2$CuO$_4$, and 6 $\times$ 6 $\times$ 6 $k$-mesh is employed for CaCuO$_2$ respectively.
To treat the screening effect accurately, we interpolate the mesh using the tetrahedron method~\cite{fujiwara03,nohara09}.
When the target band crosses the bands outside the target space,
we completely disentangle them and construct perfectly orthogonalized two separated Hilbert spaces~\cite{miyake09}.
In this procedure, we discard the off-diagonal part of the self-energy between the target and the other bands. It is justified when the off-diagonal contribution is small and this is indeed the case as is already discussed in Ref.~\onlinecite{hirayama18}.
In this way, we calculate the polarization without any arbitrariness.
We check the convergence with respect to the $k$-mesh
by comparing the result with the 8 $\times$ 8 $k$-mesh for the $ab$-directions.
By comparing the calculations with the smaller $k$-mesh, we checked that these conditions give well converged results.
We include bands about from $-25$ eV to $120$ eV for calculation of the screened interaction and the self-energy.
For entangled bands, we disentangle the target bands from the global Kohn-Sham bands~\cite{miyake09}. 
Other computational conditions are the same as those in Ref.~\onlinecite{hirayama18}.

We expect that the difference arising from the choice of basis functions (for instance, plane wave basis or localized basis) in the DFT calculation is small as was shown in a previous work~\cite{miyake10}.
However, the $3d$ orbital of Cu is relatively localized among that of the transition metals.
Therefore, the bare Coulomb interaction $v$ and the screened Coulomb interaction calculated from $v$ are sensitive to the accuracy of the wave function near the core.
When one wishes to calculate with a plane wave basis, we need more careful analyses but the accuracy of interaction may be improved by using hard pseudo potentials.

\subsection{Method and Conditions for VMC}\label{VMCmethod}
To analyze the ground states of the obtained
	low-energy effective Hamiltonians of Ag-based oxides,
	we use the many-variable VMC (mVMC) method~\cite{TaharaVMC},
	which is implemented in open-source software package ``mVMC"~\cite{mVMC_CPC,mVMC}.
	The variational wave function used in this study is
	defined as
	\begin{eqnarray}
	|\psi\ra =\sP_{\rm G}\sP_{\rm J}\sL^{S}|\phi_{\rm pair}\ra,
	\label{mvmc_wavefunction_def}
	\end{eqnarray}
	where $\sP_{\rm G}$ ($\sP_{\rm J}$)
	represents the Gutzwiller factor~\cite{Gutzwiller} (Jastrow factor~\cite{Jastrow}).
	The pair product wavefuntion $|\psi\ra$  is defined as
	\begin{eqnarray}
	|\phi_{\rm pair}\ra= \Big[\sum_{i,j=1}^{\Ns}
		\sum_{\nu,\mu=1}^{N_{\rm orb}}f_{ij\nu\mu}d_{i\nu\uparrow}^{\dag}d_{j\mu\downarrow}^{\dag}\Big]^{N_{e}/2} |0 \ra,
	\end{eqnarray}
	where $f_{ij\nu\mu}$ represents the variation parameters and
	$N_{\rm s}=L\times L$ ($N_{\rm orb}$) is the number of sites (orbitals),
where $L$ is the linear dimension of the square lattice.
	By choosing the variation parameter $f_{ij\nu\mu}$ properly,
	we can express several quantum phases such as 
	magnetic ordered phases, superconducting phases, and quantum spin liquids.

In the actual calculations, 
	we impose $2\times2$ sublattice structure 
	in the variation parameters.
	The sublattice structure is shown in Fig.~\ref{dpp2x2}.
	We assume the translation symmetry beyond this supercell. 
	Thus, we have $2\times2\times N_{\rm orb}^2 \times N_{\rm s}$
	independent variation parameters for the pair-product part.
	We also employ the total spin projection $\sL^{S}$,
	which restores the symmetry of the Hamiltonians~\cite{ring2004nuclear}.
	We use the projection into the spin singlet ($S=0$) for the total spin in the calculations.
	We optimize all the variation parameters simultaneously 
	using the stochastic reconfiguration method~\cite{Sorella_PRB2001,TaharaVMC}.
	
\mi{
To reach reliable estimates of the energy, we perform the standard method of the variance extrapolation~\cite{imadakashima,ohgoe20} by supplementing the
	 Lanczos and restricted Boltzmann machine calculations~\cite{nomura}. }
	 
\mi{
Measured physical quantities are the spin structure factor $S_s(\bm{q})$ with momentum $\bm{q}$ dependence and superconducting correlation $P_d(\bm{r})$ at spatial distance $\bm{r}$ in addition to the energy. $S_s$ is defined by
\begin{equation}
S_s({\bm q}) = \frac{1}{N_s} \sum_{i,j}^{N_s} \langle {\bm S}_i \cdot {\bm S}_j \rangle e^{i {\bm q}\cdot  ({\bm r}_i-{\bm r}_j)},
\label{Sq}
\end{equation}
where ${\bm S}_i$  is the spin-1/2 operator at the site $i$.
The magnetic ordered moment is calculated from $S_s$ as 
\begin{equation}
m=2\sqrt{S(\bm{Q})/N_s}.
\label{m}
\end{equation}
The momentum distribution is defined by 
\begin{equation}
n({\bm k}) = \frac{1}{ 2N_{\rm orb}N_s} \sum_{i,j,\ell\sigma} \langle d_{i\ell\sigma}^{\dagger} d_{j\ell\sigma} \rangle e^{i {\bm k}\cdot  ({\bm r}_i-{\bm r}_j)},
\label{n_k}
\end{equation}
where $N_{orb}$ is the number of orbitals and the orbital indices $\ell$ and its summation are not necessary for the single-band ($N_{\rm orb}$=1) Hamiltonian. 
}

\section{Structural parameters of Au-based oxides}
\label{sec:structure_params_Au}

The structure parameters of the designed Au-based oxides optimized by the PBEsol function are summarized in Tables \ref{table:structures_Tprime_Au} and \ref{table:structures_T_Au}.

\section{Band structure on the LDA level}
\label{BandLDA}

\begin{figure*}[htp]
	\includegraphics[clip,width=0.7\textwidth ]{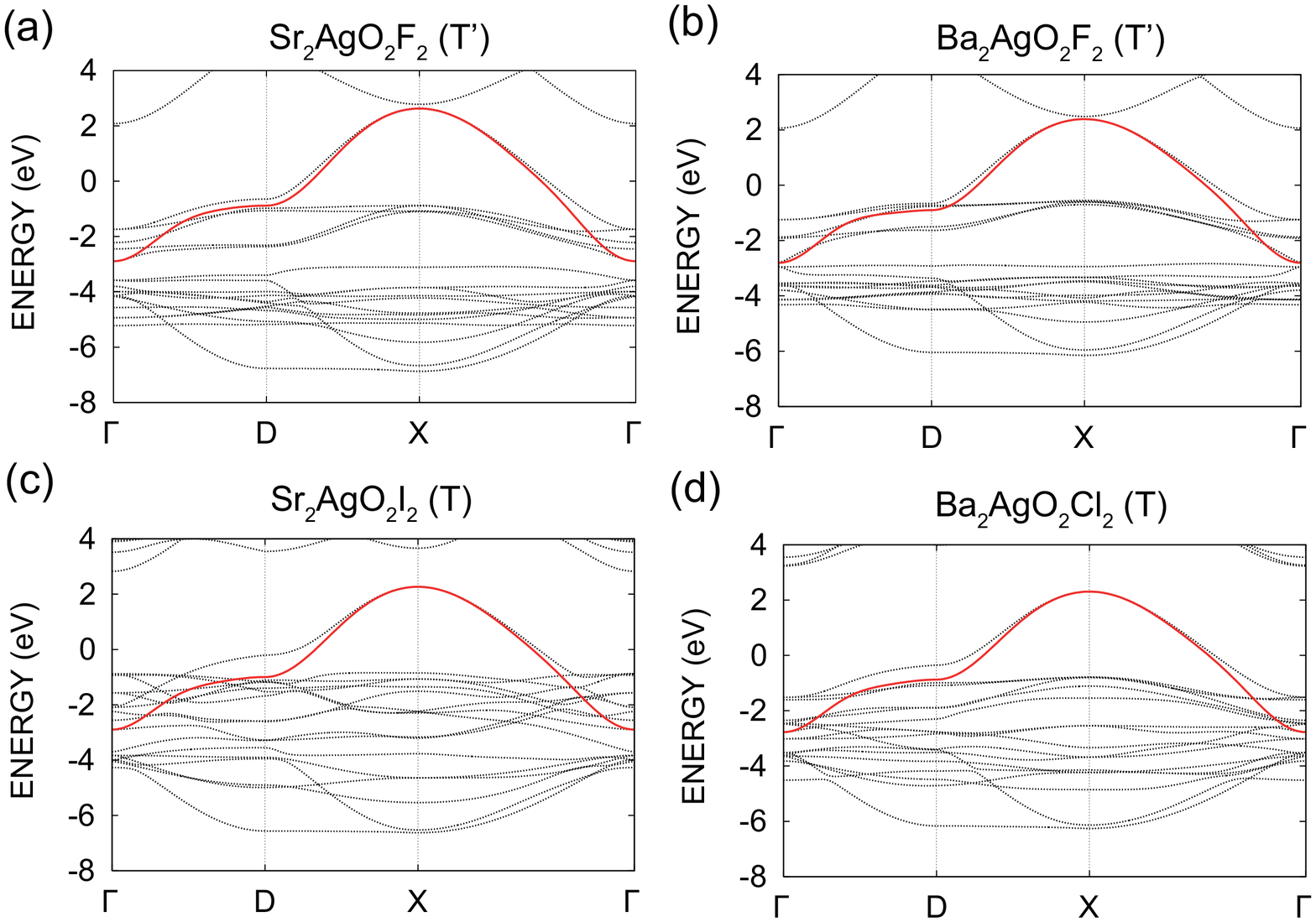}
	\caption{(Color online) 
		Electronic band structure of one-band Hamiltonian in the LDA originating from the Ag $d_{x^2-y^2}$ anti-bonding Wannier orbital.
		The zero energy corresponds to the Fermi level. 
		For comparison, the band structures in the LDA is also given (black dotted line).
	}
	\label{bndAgLDAwan1}
\end{figure*}
\begin{figure}[h!]
	\centering 
	\includegraphics[clip,width=0.45\textwidth ]{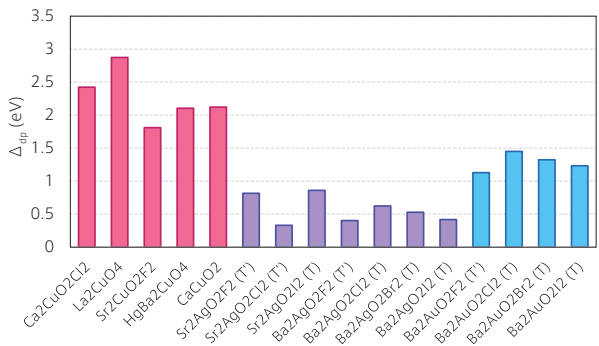} 
	\caption{(Color online) Difference of the onsite potential between the $x^2-y^2$ and the $2p$ orbitals.
	}
	\label{dp_onsite_LDA}
\end{figure} 
\begin{figure*}[htp]
	\includegraphics[clip,width=0.7\textwidth ]{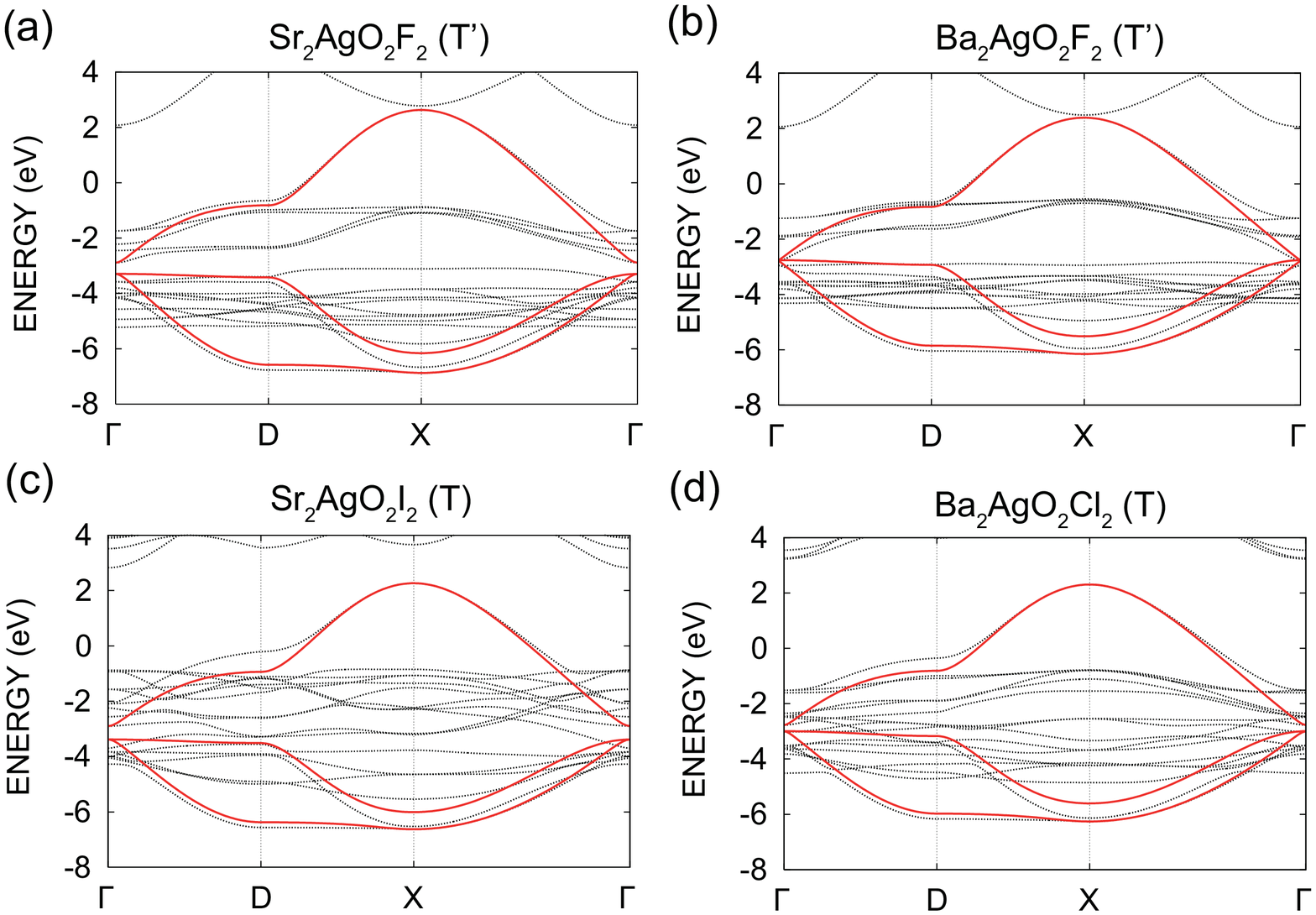}
	\caption{(Color online) 
		Electronic band structure of three-band $dpp$ Hamiltonian in the LDA originating from the Ag $d_{x^2-y^2}$ and the O $2p$ Wannier orbitals.
		The zero energy corresponds to the Fermi level. 
		For comparison, the band structures in the LDA is also given (black dotted line).
	}
	\label{bndAgLDAwan3}
\end{figure*}
\begin{figure}[h]
	\centering 
	\includegraphics[clip,width=0.45\textwidth ]{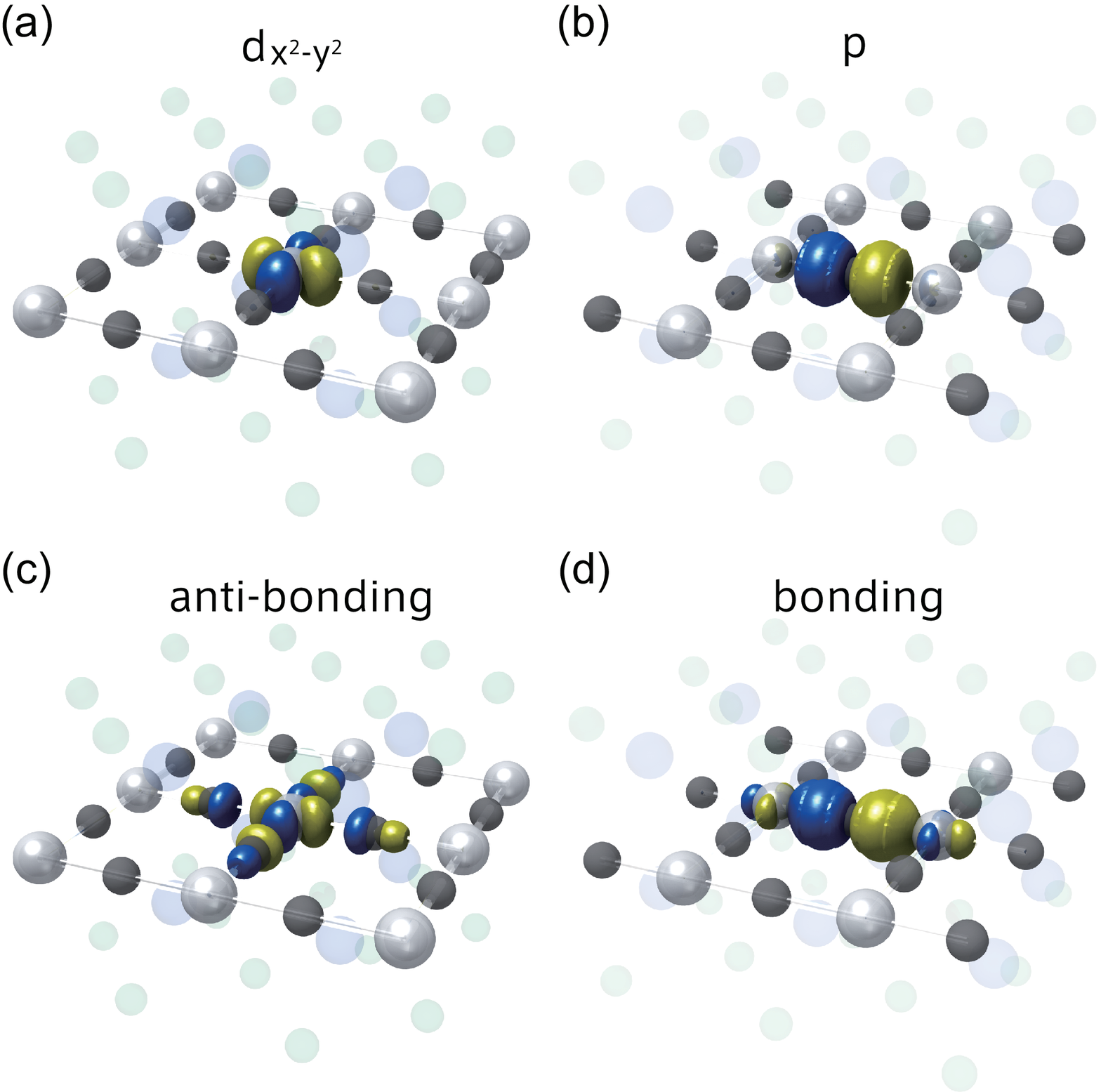} 
	\caption{(Color online) 
		Isosurface of the maximally localized Wannier function for $\pm 0.07$ a.u for (a) the Ag $d_{x^2-y^2}$ orbital, (b) the O $p$ orbital, and (c) anti-bonding and (d) bonding orbitals consisting of the Ag $x^2-y^2$ and O $p$ orbital of Sr$_2$AgO$_2$F$_2$ (T').
	}
	\label{wanAg}
\end{figure} 
%
Figure~\ref{bndAgLDAwan1} shows the band dispersion of one-band Hamiltonians constructed from the Wannier orbital for several Ag-based oxides at the LDA level.
We show the DFT/LDA band structure as a comparison.

Next, we consider the \tor{three-band} Hamiltonian.
Figure~\ref{bndAgLDAwan3} shows the fitting of the band dispersion in the \tor{three-band} Hamiltonian for the Ag $x^2-y^2$ and the O $2p$ orbitals of some Ag-based oxides.
We also show the corresponding Wannier functions of the Ag $x^2-y^2$, the O $2p$ orbitals, and the antibonding and bonding orbitals in Fig.~\ref{wanAg}.
There are three important parameters in the \tor{three-band} Hamiltonian that characterize $U/t$ of the \tor{one-band} Hamiltonian: hopping between the $x^2-y^2$ and $2p$ orbitals $t_{dp}$, the on-site potential difference between the $x^2-y^2$ and $2p$ orbitals $\Delta_{dp}$, and dielectric constant for the $x^2-y^2$ and $2p$ orbitals $U/v$, where $v$ is the bare Coulomb interaction.
Figure~\ref{dp_onsite_LDA} shows the comparison of $\Delta_{dp}$ for the copper, silver, and gold oxides.
We find that the Ag compounds are strongly covalent systems because the energies of the $x^2-y^2$ and $2p$ orbitals are very close, while the Au compounds have intermediate values of $\Delta_{dp}$.
We summarize other parameters including $\Delta_{dp}$ and $U/v$ in Tables~\ref{paraCu3_LDA},~\ref{paraAg3_LDA}, and ~\ref{paraAu3_LDA}.

\begin{figure*}[htp]
	\includegraphics[clip,width=0.9\textwidth ]{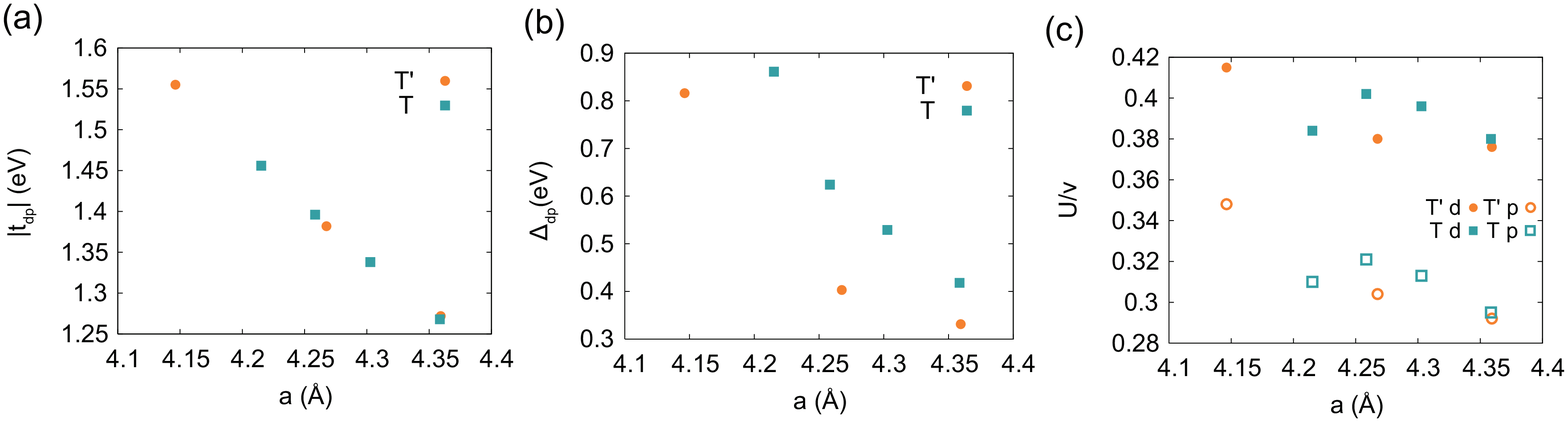}
	\caption{(Color online)
		Nearest neighbor hopping $t_{dp}$ and on-site potential difference between the Ag $x^2-y^2$ and the O $p$ orbitals, $\Delta_{dp}$ and $|U/v|$ for the Ag $x^2-y^2$ and the O $p$ orbitals plotted against the lattice constant $a$.
	}
	\label{adp}
\end{figure*}

The three parameters, $t_{dp}$, $\Delta_{dp}$ and $U/v$ show a relation with the lattice constant $a$.
Figure~\ref{adp}(a) shows hopping $t_{dp}$ versus $a$ for Ag oxides.
Regardless of the type of the structure, the hopping $t_{dp}$ decreases monotonically with increasing lattice constant $a$ as is naturally expected.
Figure~\ref{adp}(b) shows the energy difference $\Delta_{dp}$.
Due to ionic stability, Ag is mainly surrounded by anions and O is mainly surrounded by cations.
As the lattice constants become smaller, the energy difference between the $x^2-y^2$ and $2p$ orbitals also becomes larger because they are more susceptible to the electric field from them.
Although Ba$_2$AgO$_2$I$_2$ is out of trend, $U/v$ also have relations with the lattice constant $a$ (Fig.~\ref{adp}(c)).
This is because elements with smaller ionic radii are more stable in their closed-shell states, and their bands tend to appear farther away from the Fermi level, thus producing less screening effects.

\begin{table*}[h] 
\caption{
On-site potentials and effective interactions for three-band $dpp$ Hamiltonian of Cu-based compounds (in eV).
The one-body part is obtained from the fitting of the LDA band structure,
while the effective interaction is the result of the cRPA from the LDA bands.
$v$, $U$, and $J$ represent the bare Coulomb, the static values of the effective Coulomb, and exchange interactions, respectively (at $\omega=0$). 
}
\ 
\label{paraCu3_LDA} 
\begin{tabular}{c|ccc|ccc|ccc|ccc} 
\hline \hline \\ [-8pt]
 Ca$_2$CuO$_2$Cl$_2$  &       &  $t(0,0,0)$  &       &     & $v(0,0,0)$ &    &       & $U(0,0,0)$ &      &     &  $J(0,0,0)$ &    \\ [+1pt]
\hline \\ [-8pt]
&  $x^2-y^2$ &  $p_1$ &  $p_2$ & $x^2-y^2$ &  $p_1$ &  $p_2$ &  $x^2-y^2$ &  $p_1$ &  $p_2$ & $x^2-y^2$ &  $p_1$ &  $p_2$  \\ 
\hline \\ [-8pt] 
$x^2-y^2$ & -1.877 & -1.279 & 1.279   & 28.824 & 8.099 & 8.099  & 9.740&  2.613 & 2.613  &           & 0.048 & 0.048  \\ 
$p_1$         & -1.279 & -4.301 & -0.640  &  8.099 & 17.524 & 5.416 & 2.613 & 6.194&  1.727  & 0.048  &            & 0.021 \\
$p_2$        &   1.279 &-0.640 & -4.301  & 8.099& 5.416&  17.524 &  2.613 & 1.727 &6.194  &  0.048  & 0.021  &         \\
\hline \hline \\ [-8pt]
 La$_2$CuO$_4$   &       &  $t(0,0,0)$  &       &     & $v(0,0,0)$ &    &       & $U(0,0,0)$ &      &     &  $J(0,0,0)$ &    \\ [+1pt]
\hline \\ [-8pt]
&  $x^2-y^2$ &  $p_1$ &  $p_2$ & $x^2-y^2$ &  $p_1$ &  $p_2$ &  $x^2-y^2$ &  $p_1$ &  $p_2$ & $x^2-y^2$ &  $p_1$ &  $p_2$ \\ 
\hline \\ [-8pt] 
$x^2-y^2$ & -2.052&  -1.403 & 1.403   & 28.782& 8.247& 8.246 & 8.717 &1.908& 1.908  &           & 0.049 & 0.049  \\ 
$p_1$         & -1.403& -4.926 &-0.666 & 8.247& 17.773& 5.501 & 1.908 & 5.330 &1.124  & 0.049  &           & 0.019 \\
$p_2$        & 1.403 & -0.666 & -4.926  & 8.246& 5.501& 17.773 & 1.908&  1.124 &5.330 &  0.049  & 0.019 &         \\
\hline \hline \\ [-8pt]
Sr$_2$CuO$_2$F$_2$  &       &  $t(0,0,0)$  &       &     & $v(0,0,0)$ &    &       & $U(0,0,0)$ &      &     &  $J(0,0,0)$ &    \\ [+1pt]
\hline \\ [-8pt]
&  $x^2-y^2$ &  $p_1$ &  $p_2$ & $x^2-y^2$ &  $p_1$ &  $p_2$ &  $x^2-y^2$ &  $p_1$ &  $p_2$ & $x^2-y^2$ &  $p_1$ &  $p_2$ \\ 
\hline \\ [-8pt] 
$x^2-y^2$ &-1.798 &-1.155 &1.155  & 28.649& 7.917& 7.917  & 9.067 & 2.438 &2.438  &           & 0.049 & 0.049  \\ 
$p_1$         &-1.155& -3.609& -0.633  &  7.917 &16.918 &5.278 & 2.438& 5.774& 1.672 & 0.049  &            & 0.023 \\
$p_2$        &  1.155& -0.633& -3.609  & 7.917 &5.278& 16.918 & 2.438& 1.672 &5.774 &  0.049  & 0.023  &         \\
\hline \hline \\ [-8pt]  
HgBa$_2$CuO$_4$  &       &  $t(0,0,0)$  &       &     & $v(0,0,0)$ &    &       & $U(0,0,0)$ &      &     &  $J(0,0,0)$ &    \\ [+1pt]
\hline \\ [-8pt]
&  $x^2-y^2$ &  $p_1$ &  $p_2$ & $x^2-y^2$ &  $p_1$ &  $p_2$ &  $x^2-y^2$ &  $p_1$ &  $p_2$ & $x^2-y^2$ &  $p_1$ &  $p_2$ \\ 
\hline \\ [-8pt] 
$x^2-y^2$ & -2.311 & -1.261& 1.261  &28.820 &8.010 &8.010  & 8.539 &1.844 &1.844 &           & 0.048 & 0.048  \\ 
$p_1$         & -1.261& -4.416 &-0.634  &  8.010 &17.114 &5.319 & 1.844 &5.170 &1.098  & 0.048  &           & 0.020 \\
$p_2$        &  1.261& -0.634 &-4.416  & 8.010 &5.319 &17.114 & 1.844 &1.098 &5.170  &  0.048  & 0.020  &         \\
\hline \hline \\ [-8pt]  
CaCuO$_2$  &       &  $t(0,0,0)$  &       &     & $v(0,0,0)$ &    &       & $U(0,0,0)$ &      &     &  $J(0,0,0)$ &    \\ [+1pt]
\hline \\ [-8pt] 
&  $x^2-y^2$ &  $p_1$ &  $p_2$ & $x^2-y^2$ &  $p_1$ &  $p_2$ &  $x^2-y^2$ &  $p_1$ &  $p_2$ & $x^2-y^2$ &  $p_1$ &  $p_2$ \\ 
\hline \\ [-8pt] 
$x^2-y^2$ & -1.914 & -1.295 &1.295   & 28.993 &8.120 &8.120  &8.906 &2.086& 2.086  &           & 0.047 & 0.047  \\ 
$p_1$         & -1.295 &-4.036& -0.633  & 8.120 &17.844 &5.410 &2.086& 5.883 &1.257  & 0.047  &            & 0.018 \\
$p_2$        &  1.295& -0.633 &-4.036  & 8.120 &5.410& 17.844 &  2.086 &1.257 &5.883  &  0.047  & 0.018  &         \\
\hline
\hline 
\end{tabular} 
\end{table*}

\begin{table*}[h] 
\caption{
On-site potentials and effective interactions for three-band $dpp$ Hamiltonian of Ag-based compounds (in eV).
The one-body part is obtained from the fitting of the LDA band structure,
while the effective interaction is the result of the cRPA from the LDA bands.
$v$, $U$, and $J$ represent the bare Coulomb, the static values of the effective Coulomb, and exchange interactions, respectively (at $\omega=0$). 
}
\ 
\label{paraAg3_LDA} 
\begin{tabular}{c|ccc|ccc|ccc|ccc} 
\hline \hline \\ [-8pt]
Sr$_2$AgO$_2$F$_2$ (T')  &       &  $t(0,0,0)$  &       &     & $v(0,0,0)$ &    &       & $U(0,0,0)$ &      &     &  $J(0,0,0)$ &    \\ [+1pt]
\hline \\ [-8pt]
&  $x^2-y^2$ &  $p_1$ &  $p_2$ & $x^2-y^2$ &  $p_1$ &  $p_2$ &  $x^2-y^2$ &  $p_1$ &  $p_2$ & $x^2-y^2$ &  $p_1$ &  $p_2$  \\ 
\hline \\ [-8pt] 
$x^2-y^2$ & -2.900 &-1.555 &1.554   & 19.455 &7.519& 7.519  & 8.083& 2.645 &2.645  &           & 0.064 & 0.064  \\ 
$p_1$         & -1.555& -3.716& -0.584  &  7.519 &17.012 &5.030 & 2.645 &5.913 &1.725  & 0.064  &            & 0.020 \\
$p_2$        &  1.554 &-0.584 &-3.716  & 7.519 &5.030& 17.012 & 2.645& 1.725 &5.913 &  0.064  & 0.020  &         \\
\hline \hline \\ [-8pt]
Sr$_2$AgO$_2$Cl$_2$ (T')   &       &  $t(0,0,0)$  &       &     & $v(0,0,0)$ &    &       & $U(0,0,0)$ &      &     &  $J(0,0,0)$ &    \\ [+1pt]
\hline \\ [-8pt]
&  $x^2-y^2$ &  $p_1$ &  $p_2$ & $x^2-y^2$ &  $p_1$ &  $p_2$ &  $x^2-y^2$ &  $p_1$ &  $p_2$ & $x^2-y^2$ &  $p_1$ &  $p_2$ \\ 
\hline \\ [-8pt] 
$x^2-y^2$ & -2.641& -1.272& 1.272  & 19.514 &7.164 &7.164  & 7.335 &1.818& 1.818  &           & 0.055 & 0.055  \\ 
$p_1$         & -1.272& -2.972& -0.505  & 7.164 &16.845 &4.793 & 1.818 &4.927 &1.035  & 0.055  &            & 0.017 \\
$p_2$        &  1.272& -0.505& -2.972  & 7.164 &4.793 &16.845 & 1.818 &1.035 &4.927  &  0.055  & 0.017  &         \\
\hline \hline \\ [-8pt]
Sr$_2$AgO$_2$I$_2$ (T)  &       &  $t(0,0,0)$  &       &     & $v(0,0,0)$ &    &       & $U(0,0,0)$ &      &     &  $J(0,0,0)$ &    \\ [+1pt]
\hline \\ [-8pt]
&  $x^2-y^2$ &  $p_1$ &  $p_2$ & $x^2-y^2$ &  $p_1$ &  $p_2$ &  $x^2-y^2$ &  $p_1$ &  $p_2$ & $x^2-y^2$ &  $p_1$ &  $p_2$ \\ 
\hline \\ [-8pt] 
$x^2-y^2$ & -2.888& -1.456& 1.456   & 19.483 &7.387& 7.387 & 7.479& 2.042& 2.042  &           & 0.060 & 0.060  \\ 
$p_1$         &-1.456 &-3.749 &-0.548  &  7.387 &16.935 &4.947 &2.042& 5.244 &1.251  & 0.060  &            & 0.018 \\
$p_2$        &   1.456& -0.548& -3.749  & 7.387 &4.947& 16.935 &  2.042 &1.251& 5.244  &  0.060  & 0.018  &         \\
\hline \hline \\ [-8pt]
Ba$_2$AgO$_2$F$_2$ (T')  &       &  $t(0,0,0)$  &       &     & $v(0,0,0)$ &    &       & $U(0,0,0)$ &      &     &  $J(0,0,0)$ &    \\ [+1pt]
\hline \\ [-8pt]
&  $x^2-y^2$ &  $p_1$ &  $p_2$ & $x^2-y^2$ &  $p_1$ &  $p_2$ &  $x^2-y^2$ &  $p_1$ &  $p_2$ & $x^2-y^2$ &  $p_1$ &  $p_2$  \\ 
\hline \\ [-8pt] 
$x^2-y^2$ & -2.785& -1.382& 1.382   & 19.464 &7.299 &7.299  &7.400 &2.044 &2.044  &           & 0.061 & 0.061  \\ 
$p_1$         &-1.382& -3.188 &-0.551  &  7.299& 16.646 &4.880 & 2.044 &5.066& 1.271  & 0.061  &            & 0.019 \\
$p_2$        &  1.382& -0.551& -3.188  & 7.299& 4.880 &16.646 &  2.044 &1.271 &5.066  &  0.061  & 0.019  &         \\
\hline \hline \\ [-8pt]
Ba$_2$AgO$_2$Cl$_2$ (T)   &       &  $t(0,0,0)$  &       &     & $v(0,0,0)$ &    &       & $U(0,0,0)$ &      &     &  $J(0,0,0)$ &    \\ [+1pt]
\hline \\ [-8pt]
&  $x^2-y^2$ &  $p_1$ &  $p_2$ & $x^2-y^2$ &  $p_1$ &  $p_2$ &  $x^2-y^2$ &  $p_1$ &  $p_2$ & $x^2-y^2$ &  $p_1$ &  $p_2$ \\ 
\hline \\ [-8pt] 
$x^2-y^2$ & -2.762& -1.396 &1.396   & 19.455& 7.307& 7.307  & 7.815 &2.247& 2.247  &           & 0.059 &0.059  \\ 
$p_1$         & -1.396 &-3.386 &-0.535  & 7.307 &16.794& 4.892 & 2.247 &5.395& 1.405 &0.059  &            & 0.018 \\
$p_2$        &  1.396& -0.535& -3.386  & 7.307& 4.892 &16.794 & 2.247& 1.405 &5.395  & 0.059  & 0.018  &         \\
\hline \hline \\ [-8pt]
Ba$_2$AgO$_2$Br$_2$ (T)  &       &  $t(0,0,0)$  &       &     & $v(0,0,0)$ &    &       & $U(0,0,0)$ &      &     &  $J(0,0,0)$ &    \\ [+1pt]
\hline \\ [-8pt]
&  $x^2-y^2$ &  $p_1$ &  $p_2$ & $x^2-y^2$ &  $p_1$ &  $p_2$ &  $x^2-y^2$ &  $p_1$ &  $p_2$ & $x^2-y^2$ &  $p_1$ &  $p_2$ \\ 
\hline \\ [-8pt] 
$x^2-y^2$ & -2.727& -1.338& 1.338   & 19.464 &7.226 &7.226  & 7.703 &2.112& 2.112  &           & 0.058 & 0.058  \\ 
$p_1$         & -1.338 &-3.256& -0.524  &  7.226 &16.644 &4.83 & 2.112 &5.202& 1.302  & 0.058  &            & 0.018 \\
$p_2$        &   1.338& -0.524& -3.256  & 7.226& 4.838 &16.644 &  2.112 &1.302 &5.202  &  0.058  & 0.018  &         \\
\hline \hline \\ [-8pt]  
Ba$_2$AgO$_2$I$_2$ (T)  &       &  $t(0,0,0)$  &       &     & $v(0,0,0)$ &    &       & $U(0,0,0)$ &      &     &  $J(0,0,0)$ &    \\ [+1pt]
\hline \\ [-8pt]
&  $x^2-y^2$ &  $p_1$ &  $p_2$ & $x^2-y^2$ &  $p_1$ &  $p_2$ &  $x^2-y^2$ &  $p_1$ &  $p_2$ & $x^2-y^2$ &  $p_1$ &  $p_2$ \\ 
\hline \\ [-8pt] 
$x^2-y^2$ &-2.737 &-1.268 &1.268   & 19.450 &7.131& 7.131  & 7.387& 1.866& 1.866  &           & 0.057 & 0.057  \\ 
$p_1$         &-1.268& -3.155 &-0.512  &  7.131 &16.475& 4.776 & 1.866 &4.862& 1.124  & 0.057  &            & 0.018 \\
$p_2$        &  1.268& -0.512 &-3.155  & 7.131 &4.776& 16.475 &  1.866 &1.124& 4.862 &  0.057  & 0.018  &         \\
\hline
\hline 
\end{tabular} 
\end{table*} 

\begin{table*}[h] 
\caption{
On-site potentials and effective interactions for three-band $dpp$ Hamiltonian of Au-based compounds (in eV).
The one-body part is obtained from the fitting of the LDA band structure,
while the effective interaction is the result of the cRPA from the LDA bands.
$v$, $U$, and $J$ represent the bare Coulomb, the static values of the effective Coulomb, and exchange interactions, respectively (at $\omega=0$). 
}
\ 
\label{paraAu3_LDA} 
\begin{tabular}{c|ccc|ccc|ccc|ccc} 
\hline \hline \\ [-8pt]
Ba$_2$AuO$_2$F$_2$ (T')  &       &  $t(0,0,0)$  &       &     & $v(0,0,0)$ &    &       & $U(0,0,0)$ &      &     &  $J(0,0,0)$ &    \\ [+1pt]
\hline \\ [-8pt]
&  $x^2-y^2$ &  $p_1$ &  $p_2$ & $x^2-y^2$ &  $p_1$ &  $p_2$ &  $x^2-y^2$ &  $p_1$ &  $p_2$ & $x^2-y^2$ &  $p_1$ &  $p_2$  \\ 
\hline \\ [-8pt] 
$x^2-y^2$ & -2.811 &-1.591 &1.591   & 16.628& 7.282& 7.282  & 6.414& 1.968 &1.968  &           & 0.092 & 0.092  \\ 
$p_1$         &-1.591& -3.940 &-0.701  &  7.282 &15.835& 4.899 & 1.968 &4.587 &1.230  & 0.092  &            & 0.027 \\
$p_2$        &  1.591 &-0.701 &-3.940  & 7.282& 4.899 &15.835 &  1.968& 1.230& 4.587  &  0.092  & 0.027  &         \\
\hline \hline \\ [-8pt]
Ba$_2$AuO$_2$Cl$_2$ (T)   &       &  $t(0,0,0)$  &       &     & $v(0,0,0)$ &    &       & $U(0,0,0)$ &      &     &  $J(0,0,0)$ &    \\ [+1pt]
\hline \\ [-8pt]
&  $x^2-y^2$ &  $p_1$ &  $p_2$ & $x^2-y^2$ &  $p_1$ &  $p_2$ &  $x^2-y^2$ &  $p_1$ &  $p_2$ & $x^2-y^2$ &  $p_1$ &  $p_2$ \\ 
\hline \\ [-8pt] 
$x^2-y^2$ & -2.772 &-1.626 &1.626   & 16.537 &7.293 &7.293  & 6.292& 1.962& 1.962  &           & 0.092 & 0.092  \\ 
$p_1$         & -1.626& -4.223& -0.696  & 7.293& 15.823 &4.922 & 1.962 &4.635 &1.228  & 0.092  &            & 0.027 \\
$p_2$        &  1.626 &-0.696 &-4.223  & 7.293& 4.922& 15.823 & 1.962 &1.228& 4.635  &  0.092  & 0.027  &         \\
\hline \hline \\ [-8pt]
Ba$_2$AuO$_2$Br$_2$ (T)  &       &  $t(0,0,0)$  &       &     &$v(0,0,0)$ &    &       & $U(0,0,0)$ &      &     &  $J(0,0,0)$ &    \\ [+1pt]
\hline \\ [-8pt]
&  $x^2-y^2$ &  $p_1$ &  $p_2$ & $x^2-y^2$ &  $p_1$ &  $p_2$ &  $x^2-y^2$ &  $p_1$ &  $p_2$ & $x^2-y^2$ &  $p_1$ &  $p_2$ \\ 
\hline \\ [-8pt] 
$x^2-y^2$ & -2.740 &-1.564 &1.564   & 16.553 &7.227 &7.227  & 6.160 &1.799 &1.799  &           & 0.091 & 0.091  \\ 
$p_1$         & -1.564 &-4.064& -0.684  & 7.227& 15.693 &4.876 &1.799& 4.409& 1.098  & 0.091  &            & 0.027 \\
$p_2$        &   1.564& -0.684& -4.064  & 7.227& 4.876& 15.693 &  1.799& 1.098& 4.409  &  0.091  & 0.027  &         \\
\hline \hline \\ [-8pt]  
Ba$_2$AuO$_2$I$_2$ (T)  &       &  $t(0,0,0)$  &       &     & $v(0,0,0)$ &    &       & $U(0,0,0)$ &      &     &  $J(0,0,0)$ &    \\ [+1pt]
\hline \\ [-8pt]
&  $x^2-y^2$ &  $p_1$ &  $p_2$ & $x^2-y^2$ &  $p_1$ &  $p_2$ &  $x^2-y^2$ &  $p_1$ &  $p_2$ & $x^2-y^2$ &  $p_1$ &  $p_2$ \\ 
\hline \\ [-8pt] 
$x^2-y^2$ & -2.722 &-1.483& 1.483   & 16.557 &7.137 &7.137  & 6.037 &1.688& 1.688  &           & 0.089 & 0.089  \\ 
$p_1$         & -1.483 &-3.955 &-0.667  & 7.137 &15.531 &4.815 & 1.688 &4.234& 1.018  & 0.089  &            & 0.027 \\
$p_2$        &  1.483& -0.667 &-3.955  &7.137& 4.815 &15.531 &  1.688 &1.018 &4.234  &  0.089  & 0.027  &         \\
\hline
\hline 
\end{tabular} 
\end{table*} 

\section{Band structure and effective Hamiltonians on the GWA level}
\label{BandGWA}

\begin{figure*}[htp]
	\includegraphics[clip,width=0.7\textwidth ]{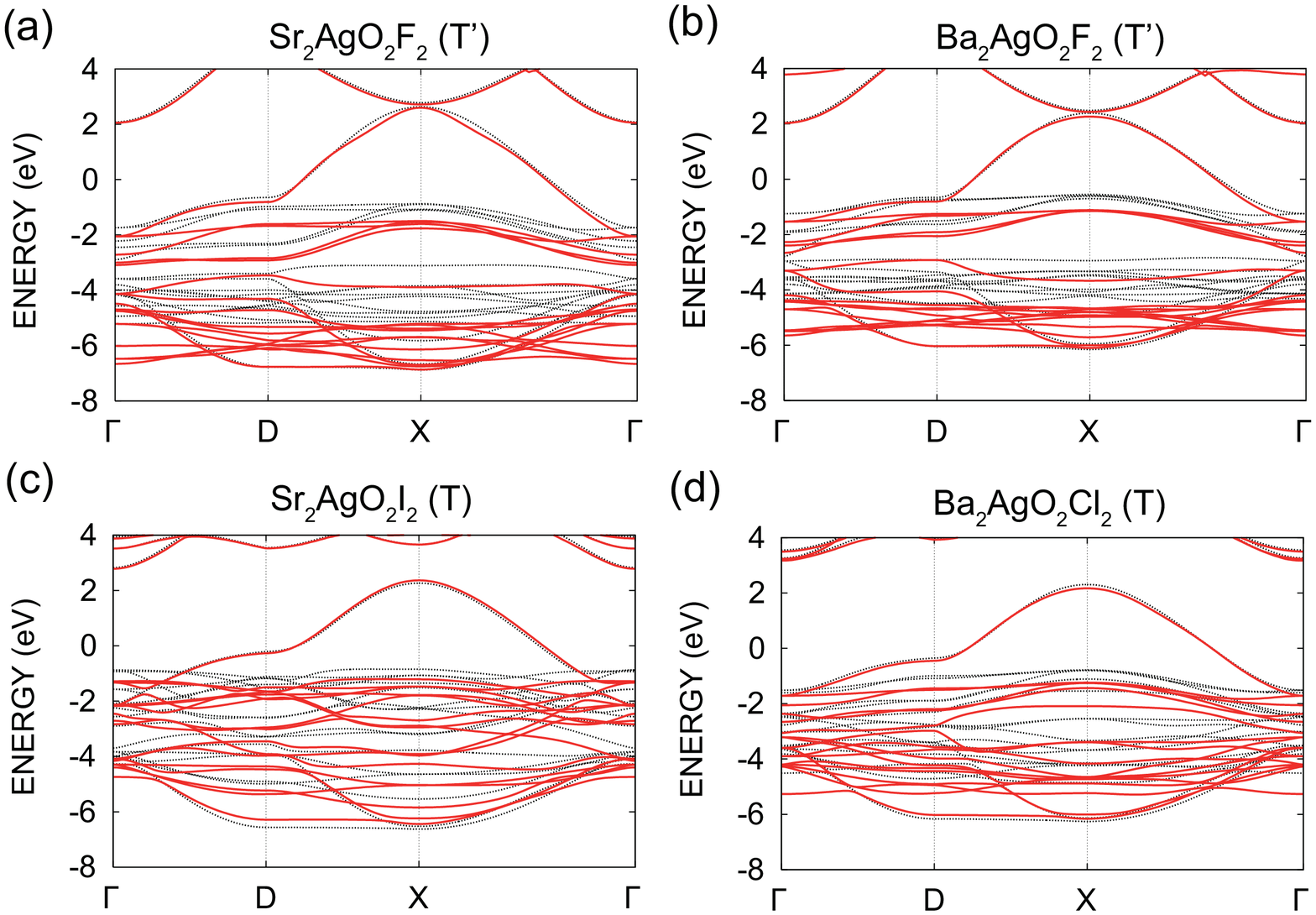}
	\caption{(Color online) 
		Electronic band structures of Ag-based compounds obtained by the GWA illustrated as red curves.
		Self-energy is calculated only for 
		the LDA 17 bands originating from the Ag $4d$ and O $2p$ orbitals near the Fermi level 
		indicated by gray in black and white plot bands.
		The zero energy corresponds to the Fermi level. 
	}
	\label{bndAgGW}
\end{figure*}
\begin{figure*}[htp]
	\includegraphics[clip,width=0.7\textwidth ]{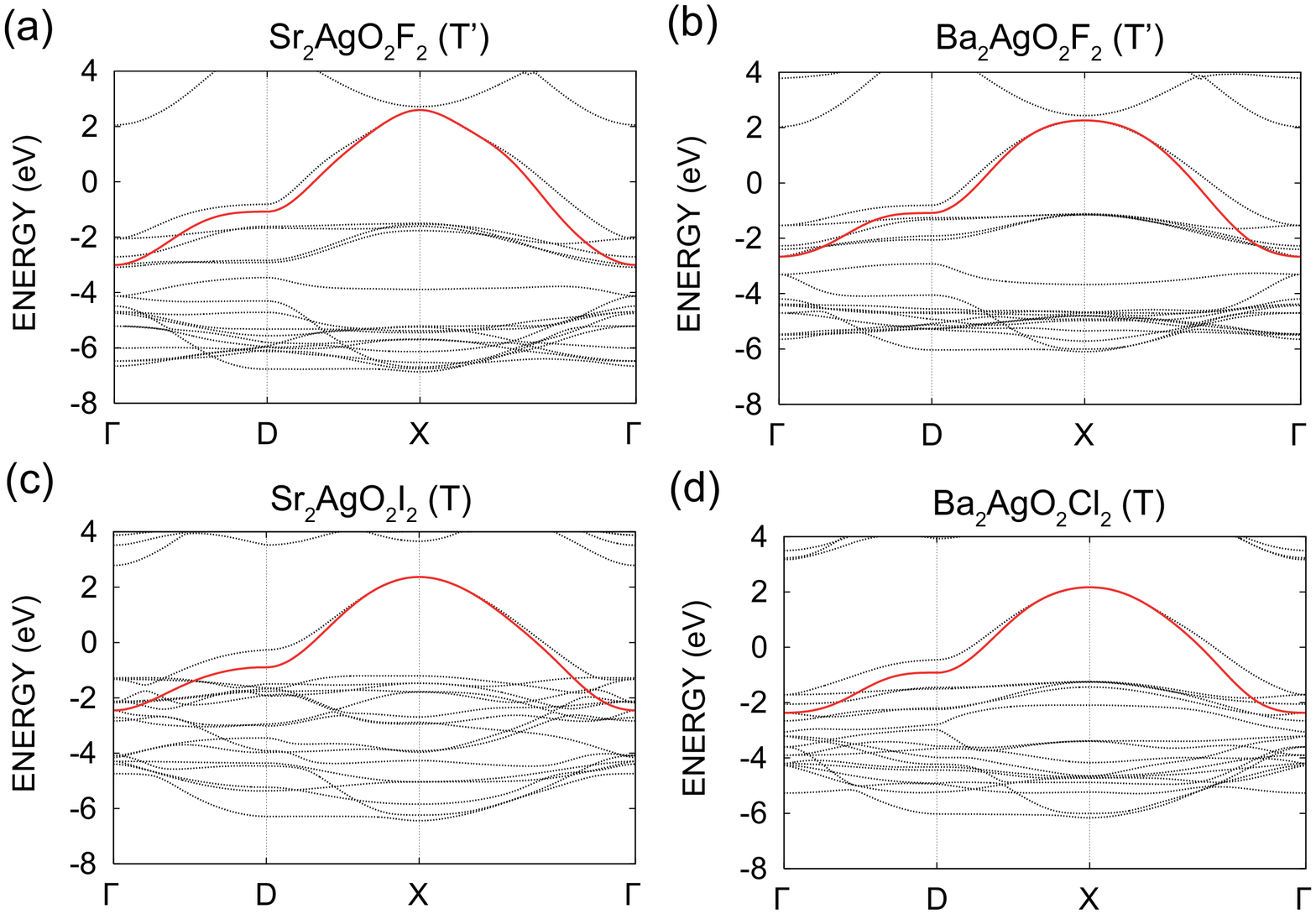}
	\caption{(Color online) 
		Electronic band structure of one-band Hamiltonian in the GWA originating from the Ag $d_{x^2-y^2}$ anti-bonding Wannier orbital.
		The zero energy corresponds to the Fermi level. 
		For comparison, the band structures in the GWA is also given (black dotted line).
	}
	\label{bndAgGWwan1}
\end{figure*}
\begin{figure*}[htp]
	\includegraphics[clip,width=0.7\textwidth ]{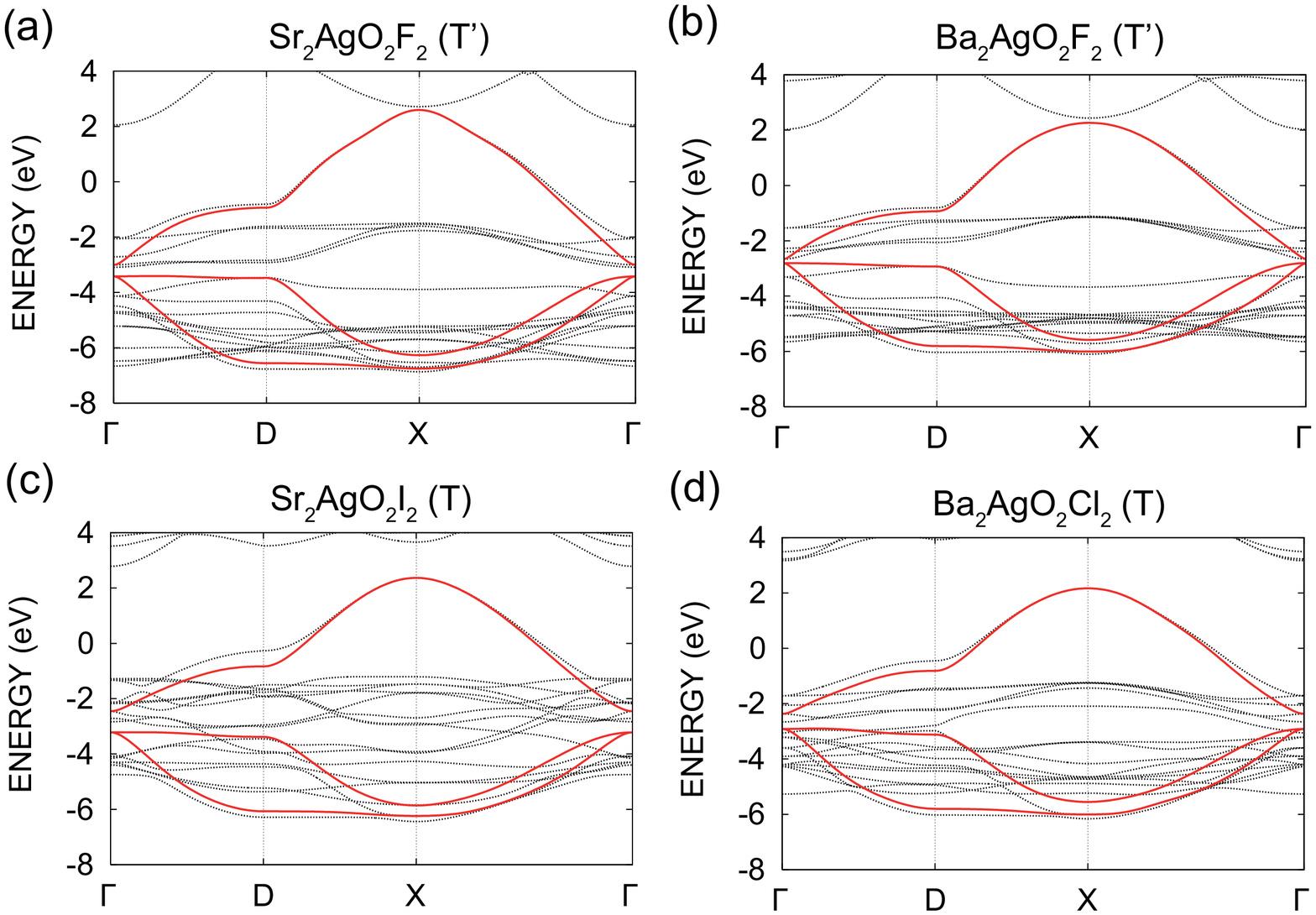}
	\caption{(Color online) 
		Electronic band structure of three-band $dpp$ Hamiltonian in the GWA originating from the Ag $d_{x^2-y^2}$ and the O $2p$ Wannier orbitals.
		The zero energy corresponds to the Fermi level. 
		For comparison, the band structures in the GWA is also given (black dotted line).
	}
	\label{bndAgGWwan3}
\end{figure*}

Figure~\ref{bndAgGW} shows the band structures in the GWA.
The band widths of the antibonding orbitals are approximately unchanged from the LDA.
This is because silver atoms are weakly correlated compared to the $3d$ elements such as copper atoms, and therefore the effect of self-energy correction of the GWA is small.
Halogen, on the other hand, has a small bandwidth and is shifted away from the Fermi level in the GWA.
Such a change has a significant effect on the screening effects.
Band structures for the one-band and three-band $dpp$ Hamiltonians on the GW level are also shown in Fig.~\ref{bndAgGWwan1} and \ref{bndAgGWwan3}, respectively.

\subsection{Downfolding}

Here, we derive low-energy effective Hamiltonians from the GW band structure.
\subsubsection{Derivation of one-band Hamiltonian}
First, as in the LDA calculations, we summarize the parameters of the \tor{one-band} Hamiltonian, the hopping calculated from the maximally localized Wannier orbitals constructed from the GW band and the effective interaction obtained by the cRPA, in Table~\ref{para1_GW}.
The $U/t$ values are larger than that in the LDA due to the self-energy effect, with the largest value being about 6.5.

\subsubsection{Derivation of three-band $dpp$ Hamiltonian}
We also summarize the \tor{three-band} Hamiltonian in Table~\ref{paraAg3_GW}.
The difference in the on-site potential between the Ag $x^2-y^2$ and O $2p$ orbitals $\Delta _{dp}$ is almost unchanged from that in the LDA.
This is because, unlike copper oxides, Ag does not have a significantly larger Coulomb interaction than O.
On the other hand, $U/v$ increases significantly because of the reduction of the screening effects owing to poor polarization by the orbitals with small bandwidth such as halogens.

\begin{table*}[h!] 
\caption{
On-site potentials and effective interactions for one-band Hamiltonian of 
Ag-based compounds (in eV).
The one-body part is obtained from the fitting of the GW band structure,
while the effective interaction is the result of the cRPA from the GW bands.
$v$, $U$, and $V$ represent the bare Coulomb, the static values of the effective on-site Coulomb, and the effective off-site Coulomb, respectively (at $\omega=0$).
$U/v$ is the ratio of the on-site bare and screened Coulomb interactions, and $U/t$ is the ratio of the nearest neighbor hopping and the screened Coulomb interaction.
}
\ 
\label{para1_GW} 
\begin{tabular}{c|c|c|c|cc|cc|cc|c|c} 
\hline \hline \\ [-8pt]
 & $t(1,0,0)$ &  $t(1,1,0)$  & $t(2,0,0)$ &  $v(0,0,0)$   & $U(0,0,0)$ & $v(1,0,0)$  & $V(1,0,0)$ & $v(1,1,0)$ & $V(1,1,0)$    &  $U/v$   &  $U/t$    \\ [+1pt]
\hline \\ [-8pt]
Sr$_2$AgO$_2$F$_2$ (T')   &  -0.682 &0.088 & -0.094  & 11.964& 4.401&  3.982 &  1.464 & 2.603 &0.966  &  0.368  &6.45         \\
\hline \\ [-8pt] 
Ba$_2$AgO$_2$F$_2$ (T')  &  -0.661 &0.100 & -0.091 &11.485 &4.010 &  3.894 & 1.282 & 2.533 &0.811  &  0.349 &6.07         \\
\hline \\ [-8pt] 
Sr$_2$AgO$_2$I$_2$ (T)  &   -0.595 &0.111 & -0.094  &11.063& 3.337& 3.907 & 0.938 & 2.564 &0.535  & 0.302  & 5.61          \\
\hline \\ [-8pt] 
Ba$_2$AgO$_2$Cl$_2$ (T)  &   -0.600 &0.079& -0.081 & 11.434 & 3.728& 3.881 &1.115 &2.536 &0.694  & 0.326 & 6.21         \\
\hline
\hline 
\end{tabular} 
\end{table*} 

\begin{table*}[h] 
\caption{
On-site potentials and effective interactions for three-band Hamiltonian of
Ag-based compounds (in eV).
The one-body part is obtained from the fitting of the GW band structure,
while the effective interaction is the result of the cRPA form the GW bands.
$v$, $U$, and $J$ represent the bare Coulomb, the static values of the effective Coulomb, and exchange interactions, respectively (at $\omega=0$). 
}
\ 
\label{paraAg3_GW} 
 \begin{tabular}{c|ccc|ccc|ccc|ccc} 
  \hline \hline \\ [-8pt]
  Sr$_2$AgO$_2$F$_2$ (T')  &       &  $t(0,0,0)$  &       &     & $v(0,0,0)$ &    &       & $U(0,0,0)$ &      &     &  $J(0,0,0)$ &    \\ [+1pt]
  \hline \\ [-8pt]
  &  $x^2-y^2$ &  $p_1$ &  $p_2$ & $x^2-y^2$ &  $p_1$ &  $p_2$ &  $x^2-y^2$ &  $p_1$ &  $p_2$ & $x^2-y^2$ &  $p_1$ &  $p_2$  \\ 
  \hline \\ [-8pt] 
  $x^2-y^2$ & -2.992& -1.534& 1.534   & 19.455& 7.519 &7.519 & 8.493& 2.865& 2.865  &           & 0.064 & 0.064  \\ 
  $p_1$         &-1.534& -3.783 &-0.607  &  7.519 &17.012 &5.030 &2.865 &6.213& 1.891 & 0.064  &            & 0.020 \\
  $p_2$        &   1.534 &-0.607 &-3.783  & 7.519 &5.030& 17.012 &  2.865& 1.891& 6.213  &  0.064  & 0.020  &         \\
  \hline \hline \\ [-8pt]
  Ba$_2$AgO$_2$F$_2$ (T')  &       &  $t(0,0,0)$  &       &     & $v(0,0,0)$ &    &       & $U(0,0,0)$ &      &     &  $J(0,0,0)$ &    \\ [+1pt]
  \hline \\ [-8pt]
  &  $x^2-y^2$ &  $p_1$ &  $p_2$ & $x^2-y^2$ &  $p_1$ &  $p_2$ &  $x^2-y^2$ &  $p_1$ &  $p_2$ & $x^2-y^2$ &  $p_1$ &  $p_2$ \\ 
  \hline \\ [-8pt] 
  $x^2-y^2$ & -2.846& -1.361& 1.361   & 19.464 &7.299& 7.299 & 8.399 &2.652& 2.652  &           & 0.061 & 0.061  \\ 
  $p_1$         & -1.361 &-3.186& -0.578  & 7.299 &16.646 &4.880 &2.652& 5.834 &1.716  & 0.061  &            & 0.019 \\
  $p_2$        &  1.361& -0.578& -3.186  & 7.299& 4.880& 16.646 &  2.652& 1.716 &5.834  &  0.061  & 0.019  &         \\
  \hline \hline \\ [-8pt]
  Sr$_2$AgO$_2$I$_2$ (T)  &       &  $t(0,0,0)$  &       &     & $v(0,0,0)$ &    &       & $U(0,0,0)$ &      &     &  $J(0,0,0)$ &    \\ [+1pt]
  \hline \\ [-8pt]
  &  $x^2-y^2$ &  $p_1$ &  $p_2$ & $x^2-y^2$ &  $p_1$ &  $p_2$ &  $x^2-y^2$ &  $p_1$ &  $p_2$ & $x^2-y^2$ &  $p_1$ &  $p_2$  \\ 
  \hline \\ [-8pt] 
  $x^2-y^2$ & -2.721 &-1.413& 1.413   & 19.483 &7.387 &7.387  & 7.672& 2.144 &2.144  &           & 0.060 & 0.060  \\ 
  $p_1$         & -1.413 &-3.568& -0.560  & 7.387 &16.935& 4.947 & 2.144& 5.370 &1.320 & 0.060  &            & 0.018 \\
  $p_2$        &   1.413& -0.560 &-3.568  & 7.387& 4.947 &16.935 & 2.144& 1.320 &5.370  &  0.060  & 0.028  &         \\
  \hline \hline \\ [-8pt]
  Ba$_2$AgO$_2$Cl$_2$ (T)   &       &  $t(0,0,0)$  &       &     & $v(0,0,0)$ &    &       & $U(0,0,0)$ &      &     &  $J(0,0,0)$ &    \\ [+1pt]
  \hline \\ [-8pt]
  &  $x^2-y^2$ &  $p_1$ &  $p_2$ & $x^2-y^2$ &  $p_1$ &  $p_2$ &  $x^2-y^2$ &  $p_1$ &  $p_2$ & $x^2-y^2$ &  $p_1$ &  $p_2$ \\ 
  \hline \\ [-8pt] 
  $x^2-y^2$ & -2.712 &-1.351 & 1.351   & 19.455& 7.307& 7.307  & 7.971 &2.321 &2.321  &           &0.059 & 0.059 \\ 
  $p_1$         & -1.351 &-3.301 &-0.544  & 7.307 &16.794 &4.892 &2.321& 5.501& 1.457 &0.059 &            & 0.018 \\
  $p_2$        &  1.351& -0.544& -3.301  &7.307 &4.892 &16.794 & 2.321 &1.457 &5.501  &  0.059  & 0.018  &         \\
  \hline
  \hline 
 \end{tabular} 
\end{table*} 

\section{Three-band $dpp$ effective Hamiltonians at the GW level for four Ag-based compounds}
\label{dpp_cGW-SIC}
\begin{figure*}[h]
	\includegraphics[clip,width=0.7\textwidth ]{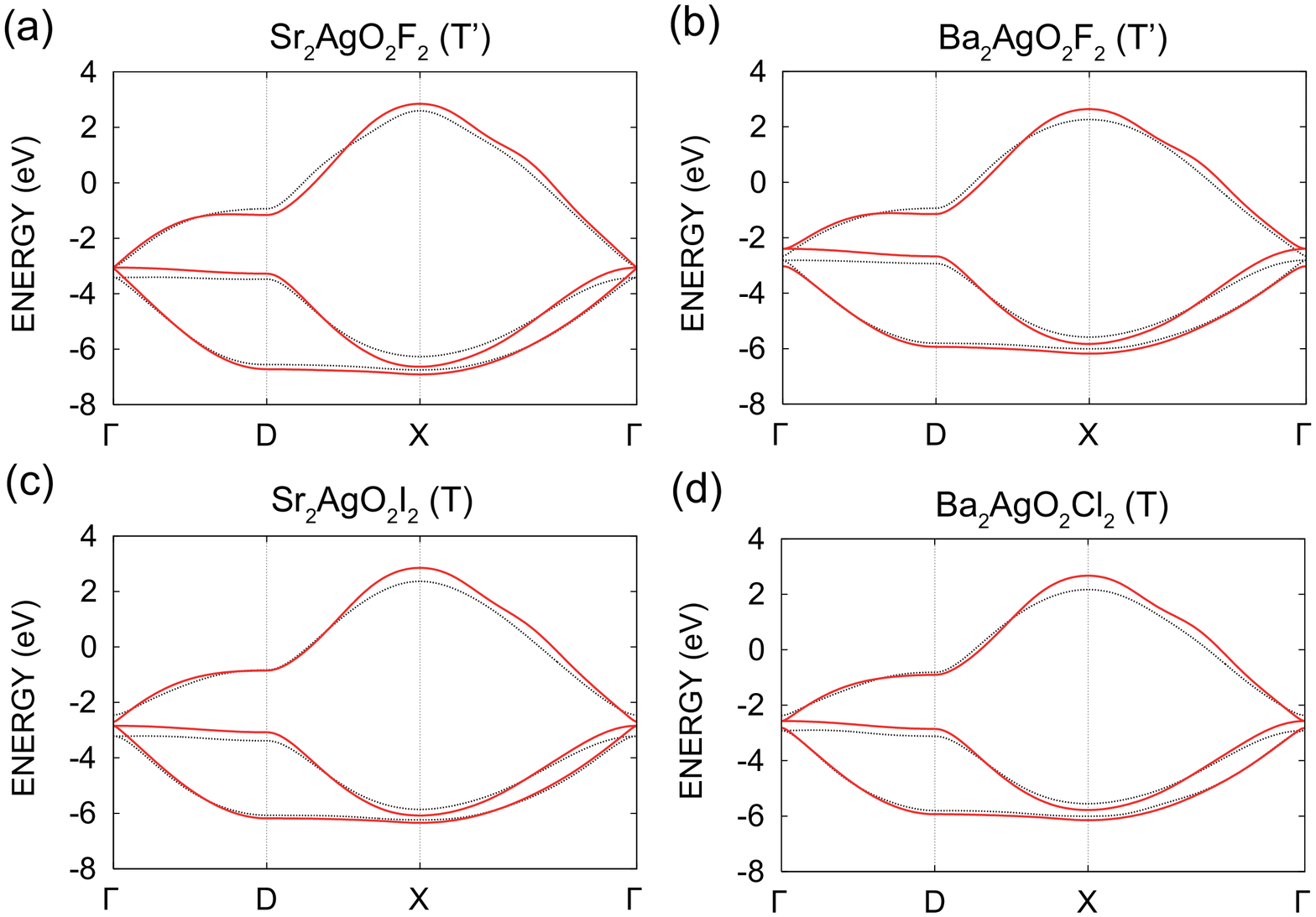}
	\caption{(Color online) 
		Electronic band structure of three-band $dpp$ Hamiltonian at the cGW-SIC level originating from the Ag $d_{x^2-y^2}$ and the O $2p$ Wannier orbitals for the four Ag compounds.
		The zero energy corresponds to the Fermi level. 
		For comparison, the fitting band structure for the GWA is also given (black dotted line).
	}
	\label{bndAgcGW-SICwan3}
\end{figure*}

Table~\ref{paraAg3_cGW-SIC} shows the parameters of the \tor{three-band} Hamiltonian calculated at the cGW-SIC level for the four Ag compounds. 
The two-body terms are the same as the GW level (see Table~\ref{paraAg3_GW} shown in Appendix~\ref{BandGWA}).
We illustrate the corresponding band structure of the one-body term in Fig.~\ref{bndAgcGW-SICwan3}.
The band width is close to that of the LDA level.
This is because the double counting of low-energy degrees of freedom to be subtracted and the renormalization arising from the frequency dependence of the interaction more or less cancel out, as is also seen in SrVO$_3$ and others~\cite{hirayama17}.
However, the energy difference between the $x^2-y^2$ and O $2p$ orbitals $\Delta _{dp}$ and hopping substantially change from those of the fitting, which have a significant impact on the correlation of the system and the stability of the superconductivity.

\begin{table*}[h] 
\caption{
On-site potentials and effective interactions for three-band $dpp$ Hamiltonian of 
Ag-based compounds (in eV).
The transfer integral is the result of the cGW-SIC from the GW bands,
while the effective interaction is the result of the cRPA from the GW bands.
$v$, $U$, and $J$ represent the bare Coulomb, the static values of the effective Coulomb, and exchange interactions, respectively (at $\omega=0$). 
}
\ 
\label{paraAg3_cGW-SIC} 
\begin{tabular}{c|ccc|ccc|ccc|ccc} 
\hline \hline \\ [-8pt]
Sr$_2$AgO$_2$F$_2$ (T')  &       &  $t(0,0,0)$  &       &     & $v(0,0,0)$ &    &       & $U(0,0,0)$ &      &     &  $J(0,0,0)$ &    \\ [+1pt]
\hline \\ [-8pt]
&  $x^2-y^2$ &  $p_1$ &  $p_2$ & $x^2-y^2$ &  $p_1$ &  $p_2$ &  $x^2-y^2$ &  $p_1$ &  $p_2$ & $x^2-y^2$ &  $p_1$ &  $p_2$  \\ 
\hline \\ [-8pt] 
$x^2-y^2$ & -3.091& -1.562& 1.562   & 19.455& 7.519 &7.519 & 8.493& 2.865& 2.865  &           & 0.064 & 0.064  \\ 
$p_1$         &-1.562& -3.702 &-0.688  &  7.519 &17.012 &5.030 &2.865 &6.213& 1.891 & 0.064  &            & 0.020 \\
$p_2$        &  1.562& -0.688& -3.702  & 7.519 &5.030& 17.012 &  2.865& 1.891& 6.213  &  0.064  & 0.020  &         \\
\hline \hline \\ [-8pt]
Ba$_2$AgO$_2$F$_2$ (T')  &       &  $t(0,0,0)$  &       &     & $v(0,0,0)$ &    &       & $U(0,0,0)$ &      &     &  $J(0,0,0)$ &    \\ [+1pt]
\hline \\ [-8pt]
&  $x^2-y^2$ &  $p_1$ &  $p_2$ & $x^2-y^2$ &  $p_1$ &  $p_2$ &  $x^2-y^2$ &  $p_1$ &  $p_2$ & $x^2-y^2$ &  $p_1$ &  $p_2$ \\ 
\hline \\ [-8pt] 
$x^2-y^2$ & -3.034& -1.388& 1.388   & 19.464 &7.299& 7.299 & 8.399 &2.652& 2.652  &           & 0.061 & 0.061  \\ 
$p_1$         & -1.388 &-3.039 &-0.648 & 7.299 &16.646 &4.880 &2.652& 5.834 &1.716  & 0.061  &            & 0.019 \\
$p_2$        &  1.388& -0.648& -3.039  & 7.299& 4.880& 16.646 &  2.652& 1.716 &5.834  &  0.061  & 0.019  &         \\
\hline \hline \\ [-8pt]
Sr$_2$AgO$_2$I$_2$ (T)  &       &  $t(0,0,0)$  &       &     & $v(0,0,0)$ &    &       & $U(0,0,0)$ &      &     &  $J(0,0,0)$ &    \\ [+1pt]
\hline \\ [-8pt]
&  $x^2-y^2$ &  $p_1$ &  $p_2$ & $x^2-y^2$ &  $p_1$ &  $p_2$ &  $x^2-y^2$ &  $p_1$ &  $p_2$ & $x^2-y^2$ &  $p_1$ &  $p_2$  \\ 
\hline \\ [-8pt] 
$x^2-y^2$ & -2.724 &-1.465 &1.465  & 19.483 &7.387 &7.387  & 7.672& 2.144 &2.144  &           & 0.060 & 0.060  \\ 
$p_1$         & -1.465& -3.409 &-0.637 & 7.387 &16.935& 4.947 & 2.144& 5.370 &1.320 & 0.060  &            & 0.018 \\
$p_2$        &  1.465& -0.637& -3.409  & 7.387& 4.947 &16.935 & 2.144& 1.320 &5.370  &  0.060  & 0.028  &         \\
\hline \hline \\ [-8pt]
Ba$_2$AgO$_2$Cl$_2$ (T)   &       &  $t(0,0,0)$  &       &     & $v(0,0,0)$ &    &       & $U(0,0,0)$ &      &     &  $J(0,0,0)$ &    \\ [+1pt]
\hline \\ [-8pt]
&  $x^2-y^2$ &  $p_1$ &  $p_2$ & $x^2-y^2$ &  $p_1$ &  $p_2$ &  $x^2-y^2$ &  $p_1$ &  $p_2$ & $x^2-y^2$ &  $p_1$ &  $p_2$ \\ 
\hline \\ [-8pt] 
$x^2-y^2$ & -2.827& -1.400& 1.400   & 19.455& 7.307& 7.307  & 7.971 &2.321 &2.321  &           &0.059 & 0.059 \\ 
$p_1$         & -1.400 &-3.162& -0.619  & 7.307 &16.794 &4.892 &2.321& 5.501& 1.457 &0.059 &            & 0.018 \\
$p_2$        & 1.400& -0.619& -3.162  &7.307 &4.892 &16.794 & 2.321 &1.457 &5.501  &  0.059  & 0.018  &         \\
\hline
\hline 
\end{tabular} 
\end{table*}

\vskip1cm
\section{Full parameters \tor{of one-band Hamiltonian for Sr$_2$AgO$_2$F$_2$}}
\tor{We summarize the full parameters of one-band Hamiltonian for Sr$_2$AgO$_2$F$_2$ in Table \ref{SrAgOF_full}.}

\label{Full parameters}
\begin{table*}[htp] 
	\caption{
		Full parameters of transfer integral and effective interaction (in eV) in one-band Hamiltonian for Sr$_2$AgO$_2$F$_2$ used for the present calculation. The notations are the same as Table~\ref{paraAg1_cGW}.
		A part of parameters is given in Table~\ref{paraAg1_cGW}.
	}
	\ 
	\label{SrAgOF_full} 
	\scalebox{0.92}[1.0]{
	\begin{tabular}{c|c|c|c|c|c|c|c|c|c|c|} 
		\hline \hline \\ [-8pt]
$t(1,0,0)$ &  $t(1,1,0)$  & $t(2,0,0)$ &  $t(2,1,0)$   & $t(2,2,0)$ & $t(3,0,0)$  & $t(3,1,0)$ & $t(3,2,0)$ & $t(3,3,0)$     \\ [+1pt]
		\hline \\ [-8pt] 
-0.6577 &  0.0920 &  -0.1014 &  0.0171  & -0.0078 &  -0.0292 &  0.0180  &  -0.0046 & -0.0038  \\  [+1pt]
		\hline \\ [-8pt] 
$v(0,0,0)$   & $U(0,0,0)$ & $V(1,0,0)$ & $V(1,1,0)$ & $V(2,0,0)$   & $V(2,1,0)$  & $V(2,2,0)$  & $V(3,0,0)$  & $V(3,1,0)$  & $V(3,2,0)$  & $V(3,3,0)$     \\ [+1pt]
     	\hline \\ [-8pt] 
11.964  &4.401 &  1.464 &  0.966 &  0.771 &  0.683 &  0.583 &  0.640 & 0.600   & 0.544 & 0.520           \\
		\hline
		\hline 
	\end{tabular} 
}
\end{table*} 

\vskip1cm
\section{Ca-doping}
\label{Ca-doping}
\begin{figure}[h!]
	\centering 
	\includegraphics[clip,width=0.45\textwidth ]{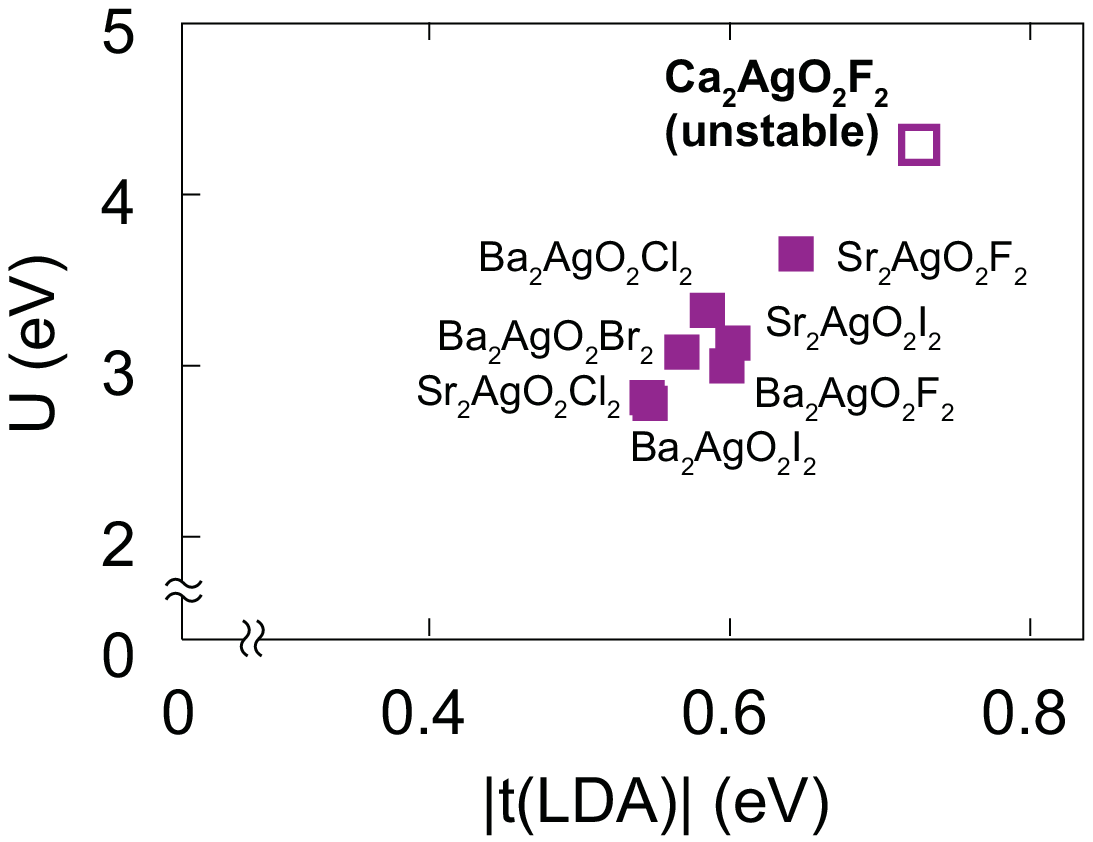}
	\caption{(Color online)
		Nearest neighbor hopping and on-site interaction in the Ag-based superconductors including unstable Ca$_2$AgO$_2$F$_2$ in the LDA.
	}
	\label{tU_LDA_Ca}
\end{figure}

Here, we show the effect of Ca-doping to alkaline earth cation.
Ca$_2$AgO$_2$F$_2$ itself is unstable, but the system with carrier doping or partial substitution for Ca might be stable.
Figure~\ref{tU_LDA_Ca} shows the nearest neighbor hopping and the on-site potential of Ca$_2$AgO$_2$F$_2$ in the LDA.
Both $t$ and $U$ become larger than those of Sr and Ba systems, while keeping the magnitude of $U/t$ approximately the same. 
We summarize the parameter for Ca$_2$AgO$_2$F$_2$ in Tables.~\ref{paraCaAg1} and ~\ref{paraCaAg3}.
\begin{table*}[h] 
	\caption{
		Transfer integral and effective interaction in one-band Hamiltonian for Ca$_2$AgO$_2$F$_2$ (T') (in eV).
		The one-body part is obtained from the fitting of the LDA band structure, 
		while the effective interaction is the result of the cRPA.
		$v$, $U$, and $V$ represent the bare Coulomb, the static values of the effective on-site Coulomb, and the effective off-site Coulomb, respectively (at $\omega=0$).
		$U/v$ is the ratio of the on-site bare and screened Coulomb interactions, and $U/t$ is the ratio of the nearest neighbor hopping and the screened Coulomb interaction. 
	}
	\ 
	\label{paraCaAg1}
		\scalebox{0.92}[1.0]{
		\begin{tabular}{c|c|c|c|cc|cc|cc|cc|c|c} 
			\hline \hline \\ [-8pt]
			LDA  & $t(1,0,0)$ &  $t(1,1,0)$  & $t(2,0,0)$ &  $v(0,0,0)$   & $U(0,0,0)$ & $v(1,0,0)$  & $V(1,0,0)$ & $v(1,1,0)$ & $V(1,1,0)$  & $v(2,0,0)$  & $V(2,0,0)$  &  $U/v$   &  $|U/t|$   \\ [+1pt]
			\hline \\ [-8pt] 
			 $x^2-y^2$  &-0.726 &  0.109 & -0.098  &  12.059  &4.291 &  4.068  & 1.410 & 2.671  & 0.915 & 2.011  & 0.717 & 0.356 & 5.91          \\
			 \hline
			 \hline 
		 \end{tabular} 
 }
\end{table*} 

\begin{table*}[h] 
	\caption{
		On-site potentials and effective interactions for three-band $dpp$ Hamiltonian of Ca$_2$AgO$_2$F$_2$ (T') (in eV).
		The one-body part is obtained from the fitting of the LDA and GW band structures and the cGW-SIC, 
		while the effective interaction is the result of the cRPA.
		$v$, $U$, and $J$ represent the bare Coulomb, the static values of the effective Coulomb, and exchange interactions, respectively (at $\omega=0$). 
	}
	\ 
	\label{paraCaAg3} 
	\begin{tabular}{c|ccc|ccc|ccc|ccc} 
		\hline \hline \\ [-8pt]
		 LDA  &       &  $t(0,0,0)$  &       &     & $v(0,0,0)$ &    &       & $U(0,0,0)$ &      &     &  $J(0,0,0)$ &    \\ [+1pt]
		\hline \\ [-8pt]
		&  $x^2-y^2$ &  $p_1$ &  $p_2$ & $x^2-y^2$ &  $p_1$ &  $p_2$ &  $x^2-y^2$ &  $p_1$ &  $p_2$ & $x^2-y^2$ &  $p_1$ &  $p_2$  \\ 
		\hline \\ [-8pt] 
		$x^2-y^2$ & -3.043 &-1.715 &1.715   & 19.392 &7.707& 7.707  & 8.167& 2.837 &2.837  &           &0.067 & 0.067  \\ 
		$p_1$     & -1.715& -4.214& -0.615  & 7.707 &17.012 &5.030 & 2.837 &6.218 &1.891  & 0.067  &            & 0.020 \\
		$p_2$     &  1.715 &-0.615 &-4.214  &7.707 &5.030& 17.012 & 2.837 & 1.891 &6.218 &  0.067  & 0.020  &         \\
		\hline \hline \\ [-8pt]
		GW   &       &  $t(0,0,0)$  &       &     & $v(0,0,0)$ &    &       & $U(0,0,0)$ &      &     &  $J(0,0,0)$ &    \\ [+1pt]
		\hline \\ [-8pt]
		&  $x^2-y^2$ &  $p_1$ &  $p_2$ & $x^2-y^2$ &  $p_1$ &  $p_2$ &  $x^2-y^2$ &  $p_1$ &  $p_2$ & $x^2-y^2$ &  $p_1$ &  $p_2$ \\ 
		\hline \\ [-8pt] 
		$x^2-y^2$ & -3.101 & -1.695& 1.695  & 19.392 &7.704 &7.704  & 8.330&2.854 &2.854  &           & 0.067 & 0.067  \\ 
		$p_1$     & -1.695& -4.287&-0.639  & 7.704 &17.300 &5.163 & 2.854 &6.252 &1.882  & 0.067  &            & 0.020 \\
		$p_2$     &  1.695& -0.639& -4.287  & 7.704 &5.163&17.300 & 2.854 &1.882 &6.252 &  0.067  & 0.020  &         \\
		\hline \hline \\ [-8pt]
		cGW-SIC  &       &  $t(0,0,0)$  &       &     & $v(0,0,0)$ &    &       & $U(0,0,0)$ &      &     &  $J(0,0,0)$ &    \\ [+1pt]
		\hline \\ [-8pt]
		&  $x^2-y^2$ &  $p_1$ &  $p_2$ & $x^2-y^2$ &  $p_1$ &  $p_2$ &  $x^2-y^2$ &  $p_1$ &  $p_2$ & $x^2-y^2$ &  $p_1$ &  $p_2$ \\ 
		\hline \\ [-8pt] 
		$x^2-y^2$ & -3.045 & -1.648& 1.648  & 19.392 &7.704 &7.704  & 8.330&2.854 &2.854  &           & 0.067 & 0.067  \\ 
		$p_1$     & -1.648& -4.047&-0.669  & 7.704 &17.300 &5.163 & 2.854 &6.252 &1.882  & 0.067  &            & 0.020 \\
		$p_2$     & 1.648& -0.669& -4.047 & 7.704 &5.163&17.300 & 2.854 &1.882 &6.252 &  0.067  & 0.020  &         \\	
		\hline \\ [-8pt]
		occ.(GWA)      & $x^2-y^2$ &  $p_1$ &  $p_2$  \\ 
		\hline \\ [-8pt] 
		& 1.518 & 1.741 & 1.741   \\
		\hline
		\hline 
	\end{tabular} 
\end{table*}


\begin{thebibliography}{99}
\bibitem{LR}
In the case of the Ag compounds, we do not consider the level renormalization taken into account for the Hg compound before~\cite{hirayama19} because the Hamiltonian and its solution seem to be insensitive to it.
\end{thebibliography}



\end{document}